\newif\ifsinglecolumn  
\singlecolumnfalse

\ifsinglecolumn
	\documentclass[10pt,onecolumn,twoside,draftclsnofoot]{IEEEtran}
\else
	\documentclass[10pt,twocolumn,twoside]{IEEEtran}
\fi

\usepackage[utf8]{inputenc}

\usepackage[usenames,dvipsnames,svgnames]{xcolor}

\usepackage{todonotes}

\usepackage[acronyms,nogroupskip,nopostdot,nonumberlist,order=word]{glossaries}
\usepackage{glossary-mcols}

\usepackage[backend=biber, natbib=true, style=ieee,doi=false,maxbibnames=5]{biblatex}

\usepackage{bm}

\usepackage{bbold}

\usepackage{accents}

\usepackage{comment}

\usepackage[nocomma,c1]{optidef}

\usepackage{caption,subcaption}
\usepackage{units}

\usepackage[autolanguage]{numprint}

\usepackage{enumitem}
\setitemize{noitemsep,topsep=0pt}

\usepackage{siunitx}

\usepackage{setspace}

\usepackage[hidelinks]{hyperref}
\usepackage[capitalize]{cleveref}

\AtEveryBibitem{\clearfield{pages}}

\glsdisablehyper 

\renewcommand{\vec}{\bm}
\newcommand{\mat}{\bm}

\renewcommand{\H}{^H}

\DeclarePairedDelimiter{\parentheses}{(}{)}

\newcommand{\diag}{\operatorname{diag}\parentheses}

\DeclarePairedDelimiter\myabs{\lvert}{\rvert}%
\DeclarePairedDelimiter\mynorm{\lVert}{\rVert}%

\makeatletter
\let\oldmyabs\myabs
\def\myabs{\@ifstar{\oldmyabs}{\oldmyabs*}}
\let\oldmynorm\mynorm
\def\mynorm{\@ifstar{\oldmynorm}{\oldmynorm*}}
\makeatother

\newcommand{\arange}[2][1]{\lbrace #1, \dots, #2 \rbrace}

\newcommand{\csum}[1]{\sum_{\scriptstyle\mathclap{#1}}}

\newcommand{\foft}[1][]{\lbrack\gls{t}#1\rbrack}

\makeatletter
\newcommand{\pushright}[1]{\ifmeasuring@#1\else\omit\hfill$\displaystyle#1$\fi\ignorespaces}
\newcommand{\pushleft}[1]{\ifmeasuring@#1\else\omit$\displaystyle#1$\hfill\fi\ignorespaces}
\makeatother

\crefname{appsec}{Appendix}{Appendices}

\newcommand{\jvspace}[1]{}

\newglossary[tlg_names]{names}{tld_names}{tdn_names}{names}
\newglossary[tlg_amod]{amod}{tld_amod}{tdn_amod}{AMoD system}
\newglossary[tlg_opf]{opf}{tld_opf}{tdn_opf}{OPF problem}

\makeglossaries

\setacronymstyle{long-short}
\newacronym{opf}{OPF}{Optimal Power Flow}
\newacronym{bfm_opf}{BFM-OPF}{branch flow model OPF}
\newacronym{bfm_sdp}{BFM-SDP}{branch flow model SDP}
\newacronym{lpfm_lp}{LPFM-LP}{linearized power flow manifold LP}
\newacronym{sdp}{SDP}{semi-definite program}
\newacronym{lp}{LP}{linear program}
\newacronym{milp}{MILP}{mixed-integer linear program}
\newacronym{qp}{QP}{quadratic program}
\newacronym{amod}{AMoD}{Autonomous Mobility on Demand}
\newacronym{pamod}{P-AMoD}{Power in the Loop Autonomous Mobility on Demand}
\newacronym{ev}{EV}{electric vehicle}
\newacronym{av}{AV}{autonomous vehicle}
\newacronym{pge}{PG\&E}{Pacific Gas and Electric Company}
\newacronym{sce}{SCE}{Southern California Edison}
\newacronym{slac}{SLAC}{SLAC National Accelerator Laboratory}
\newacronym{bfm}{BFM}{branch flow model}
\newacronym{bim}{BIM}{bus injection model}
\newacronym{iso}{ISO}{independent system operator}
\newacronym{v2g}{V2G}{vehicle-to-grid}
\newacronym{soc}{SoC}{state-of-charge}
\newacronym{bfm_lp}{BFM-LP}{branch flow model LP}
\newacronym{pdn}{PDN}{power distribution network}
\newglossaryentry{t}
{
	name={\ensuremath{t}},
	sort={t},
	description={time}
}

\newglossaryentry{t_set}
{
	type=amod,
	name={\ensuremath{\mathcal{T}}},
	sort={T},
	description={set of time steps}
}

\newglossaryentry{n_t}
{
	name={\ensuremath{T}},
	sort={T},
	description={number of time steps}
}

\newglossaryentry{t_step}
{
	type=amod,
	name={\ensuremath{\Delta\gls{t}}},
	sort={zz D t},
	description={length of a time step}
}

\newglossaryentry{road_graph}
{
	name={\ensuremath{R}},
	description={road graph}
}

\newglossaryentry{road_node_set}
{
	type=amod,
	name={\ensuremath{\mathcal{V}_{\gls{road_graph}}}},
	sort={V R},
	description={set of road vertices}
}

\newglossaryentry{road_edge_set}
{
	type=amod,
	sort={A R},
	name={\ensuremath{\mathcal{A}_{\gls{road_graph}}}},
	description={set of road arcs}
}

\newglossaryentry{in_road_node}
{
	name={\ensuremath{v}},
	description={road vertex}
}

\newglossaryentry{in_road_node_post}
{
	name={\ensuremath{w}},
	description={xx}
}

\newglossaryentry{in_road_node_pre}
{
	name={\ensuremath{u}},
	description={xx}
}

\newglossaryentry{in_road_edge}
{
	name={\ensuremath{\gls{in_road_node},\gls{in_road_node_post}}},
	description={road arc}
}

\newglossaryentry{d}
{
	name={\ensuremath{d}},
	description={distance}
}

\newglossaryentry{d_edge}
{
	type=amod,
	name={\ensuremath{\gls{d}_{\gls{in_road_edge}}}},
	sort={d v w},
	description={distance of road arc $(\gls{in_road_edge}) \in \gls{road_edge_set}$}
}

\newglossaryentry{t_edge}
{
	type=amod,
	name={\ensuremath{\gls{t}_{\gls{in_road_edge}}}},
	sort={t v w},
	description={time to traverse road arc $(\gls{in_road_edge}) \in \gls{road_edge_set}$}
}

\newglossaryentry{c}
{
	name={\ensuremath{c}},
	description={charge}
}

\newglossaryentry{n_c}
{
	name={\ensuremath{C}},
	sort={C},
	description={number of discrete battery charge levels}
}

\newglossaryentry{c_set}
{
	type=amod,
	name={\ensuremath{\mathcal{C}}},
	sort={C},
	description={set of discrete battery charge levels}
}

\newglossaryentry{c_unit}
{
	type=amod,
	name={\ensuremath{\gls{E}_{\gls{c}}}},
	sort={E c},
	description={amount of energy in a charge level}
}

\newglossaryentry{c_edge}
{
	type=amod,
	name={\ensuremath{\gls{c}_{\gls{in_road_edge}}}},
	sort={c v w},
	description={energy consumption for traversing road arc $(\gls{in_road_edge})\in\gls{road_edge_set}$}
}

\newglossaryentry{f}
{
	name={\ensuremath{f}},
	description={network flow}
}

\newglossaryentry{f_0}
{
	type=amod,
	name={\ensuremath{\gls{f}_0}},
	sort={f 0},
	description={\glsdesc{f} for rebalancing vehicles}
}

\newglossaryentry{flow_m}
{
	type=amod,
	name={\ensuremath{\gls{f}_{\gls{in_m}}}},
	sort={f m},
	description={\glsdesc{f} for customer trip request $\gls{in_m} \in \gls{m_set}$}
}

\newglossaryentry{f_cap}
{
	name={\ensuremath{\bar{\gls{f}}}},
	description={road capacity}
}

\newglossaryentry{f_edge_cap}
{
	type=amod,
	name={\ensuremath{\gls{f_cap}_{\gls{in_road_edge}}}},
	sort={f v w},
	description={maximum capacity of road arc $(\gls{in_road_edge})\in \gls{road_edge_set}$}
}

\newglossaryentry{f_exp_edge_cap_t}
{
	type=amod,
	name={\ensuremath{\ensuremath{\gls{f_cap}_{(\gls{in_road_edge}),\gls{t}}}}},
	sort={f v w},
	description={residual road capacity}
}

\newglossaryentry{in_charger}
{
	name={\ensuremath{s}},
	description={charging station}
}

\newglossaryentry{n_charger}
{
	name={\ensuremath{|\gls{charger_set}|}},
	description={xx}
}

\newglossaryentry{charger_set}
{
	type=amod,
	name={\ensuremath{\mathcal{S}}},
	sort={S},
	description={set of chargers in the road network}
}

\newglossaryentry{charge_rate_charger}
{
	type=amod,
	name={\ensuremath{\delta_{\gls{n_c},\gls{in_charger}}}},
	sort={zz d C s},
	description={charging rate of charger $\gls{in_charger}\in\gls{charger_set}$}
}

\newglossaryentry{charger_cap}
{
	type=amod,
	name={\ensuremath{\bar{S}_{\gls{in_charger}}}},
	sort={S c},
	description={number of charging plugs in \glsdesc{in_charger} $\gls{in_charger} \in \gls{charger_set}$}
}

\newglossaryentry{exp_network}
{
	name={\ensuremath{G}},
	description={expanded road network}
}

\newglossaryentry{exp_node_set}
{
	type=amod,
	name={\ensuremath{\mathcal{V}}},
	sort={V},
	description={set of expanded graph vertices}
}

\newglossaryentry{exp_edge_set}
{
	type=amod,
	name={\ensuremath{\mathcal{A}}},
	sort={A},
	description={set of expanded graph arcs}
}

\newglossaryentry{exp_edge_set_road}
{
	type=amod,
	name={\ensuremath{\gls{exp_edge_set}_{T}}},
	sort={A T},
	description={\glsdesc{exp_edge_set} representing a physical time-dependent movement in the road network}
}

\newglossaryentry{exp_edge_set_charger}
{
	type=amod,
	name={\ensuremath{\gls{exp_edge_set}_{S}}},
	sort={A S},
	description={\glsdesc{exp_edge_set} representing a recharging process}
}

\newglossaryentry{in_exp_node}
{
	name={\ensuremath{\bm{\mathrm{\gls{in_road_node}}}}},
	description={expanded network node}
}

\newglossaryentry{in_exp_node_post}
{
	name={\ensuremath{\bm{\mathrm{\gls{in_road_node_post}}}}},
	description={xx}
}

\newglossaryentry{in_exp_node_pre}
{
	name={\ensuremath{\bm{\mathrm{\gls{in_road_node_pre}}}}},
	description={xx}
}

\newglossaryentry{in_exp_edge}
{
	name={\ensuremath{\gls{in_exp_node},\gls{in_exp_node_post}}},
	description={expanded network arc}
}

\newglossaryentry{in_exp_edge_pre}
{
	name={\ensuremath{\gls{in_exp_node_pre},\gls{in_exp_node}}},
	description={expanded network arc}
}

\newglossaryentry{in_exp_node_road}
{
	type=amod,
	name={\ensuremath{\gls{in_road_node}_{\gls{in_exp_node}}}},
	sort={v v},
	description={road vertex associated to expanded vertex $\gls{in_exp_node} \in \gls{exp_node_set}$}
}

\newglossaryentry{in_exp_edge_road}
{
	name={\ensuremath{\gls{in_exp_node_road},\gls{in_exp_node_road_post}}},
	description={xx}
}

\newglossaryentry{in_exp_node_t}
{
	type=amod,
	name={\ensuremath{\gls{t}_{\gls{in_exp_node}}}},
	sort={t v},
	description={time step associated to expanded vertex $\gls{in_exp_node} \in \gls{exp_node_set}$}
}

\newglossaryentry{in_exp_node_c}
{
	type=amod,
	name={\ensuremath{\gls{c}_{\gls{in_exp_node}}}},
	sort={c v},
	description={\gls{soc} associated to expanded vertex $\gls{in_exp_node} \in \gls{exp_node_set}$}
}

\newglossaryentry{in_exp_node_road_post}
{
	name={\ensuremath{\gls{in_road_node}_{\gls{in_exp_node_post}}}},
	description={xx}
}

\newglossaryentry{in_exp_node_t_post}
{
	name={\ensuremath{\gls{t}_{\gls{in_exp_node_post}}}},
	description={xx}
}

\newglossaryentry{in_exp_node_c_post}
{
	name={\ensuremath{\gls{c}_{\gls{in_exp_node_post}}}},
	description={xx}
}

\newglossaryentry{n_m}
{
	name={\ensuremath{M}},
	sort={M},
	description={number of customer trip requests}
}

\newglossaryentry{in_m}
{
	name={\ensuremath{m}},
	description={customer trip request}
}

\newglossaryentry{m_set}
{
	type=amod,
	name={\ensuremath{\mathcal{M}}},
	sort={M},
	description={set of customer trip requests}
}

\newglossaryentry{trip_rate}
{
	name={\ensuremath{\lambda}},
	description={customer trip rate}
}

\newglossaryentry{trip_rate_m}
{
	type=amod,
	name={\ensuremath{\lambda_{\gls{in_m}}}},
	sort={zz l m},
	description={customer rate of trip request $\gls{in_m} \in \gls{m_set}$}
}

\newglossaryentry{origin_node_m}
{
	type=amod,
	name={\ensuremath{\gls{in_road_node}_{\gls{in_m}}}},
	sort={v m},
	description={origin of trip request $\gls{in_m} \in \gls{m_set}$}
}

\newglossaryentry{dest_node_m}
{
	type=amod,
	name={\ensuremath{\gls{in_road_node_post}_{\gls{in_m}}}},
	sort={w m},
	description={destination of trip request $\gls{in_m} \in \gls{m_set}$}
}

\newglossaryentry{t_m}
{
	type=amod,
	name={\ensuremath{\gls{t}_{\gls{in_m}}}},
	sort={t m},
	description={departure timestep of trip request $\gls{in_m} \in \gls{m_set}$}
}

{
\newglossaryentry{route_precomp}
{
	name={\ensuremath{r_{\gls{in_road_node}\to\gls{in_road_node_post}}}},
	sort={r v z w},
	description={precomputed customer route from $\gls{in_road_node} \in \gls{road_node_set}$ to $\gls{in_road_node_post} \in \gls{road_node_set}$}
}

\newglossaryentry{route_precomp_t}
{
	name={\ensuremath{\gls{t}_{\gls{in_road_node}\to\gls{in_road_node_post}}}},
	sort={t v z w},
	description={traveling time of \glsdesc{route_precomp}}
}

\newglossaryentry{route_precomp_c}
{
	name={\ensuremath{\gls{c}_{\gls{in_road_node}\to\gls{in_road_node_post}}}},
	sort={c v z w},
	description={energy consumption of \glsdesc{route_precomp}}
}
}

{

\newglossaryentry{price}
{
	name={\ensuremath{V}},
	description={price}
}

\newglossaryentry{distance_price}
{
	type=amod,
	name={\ensuremath{\gls{price}_{D}}},
	sort={V D},
	description={vehicle operation cost per unit distance (excluding electricity)}
}

\newglossaryentry{electricity_price}
{
	name={\ensuremath{\gls{price}_{\gls{el}}}},
	description={price of electricity}
}

\newglossaryentry{electricity_price_charger}
{
	type=amod,
	name={\ensuremath{\gls{price}_{\gls{el},\gls{in_charger}}}},
	sort={V el s},
	description={\glsdesc{electricity_price} in charging station $\gls{in_charger}\in\gls{charger_set}$}
}

\newglossaryentry{electricity_price_dist}
{
	type=amod,
	name={\ensuremath{\gls{price}_{\gls{el},\gls{in_dist}}}},
	sort={V el d},
	description={\glsdesc{electricity_price} at the substation of network $\gls{in_dist} \in \gls{in_dist_set}$}
}

\newglossaryentry{electricity_price_dist_set}
{
	type=amod,
	name={\ensuremath{\gls{price}_{\gls{el},\gls{in_dist_set}}}},
	sort={V el D},
	description={total \glsdesc{electricity_price} for all \gls{n_dist} \glsdescplural{in_dist}}
}

}

\newglossaryentry{trip_rate_in_c}
{
	type=amod,
	name={\ensuremath{\gls{trip_rate_m}^{\gls{c},\gls{in}}}},
	description={number of vehicles with \glsdesc{c} \gls{c} departing to serve \glsdesc{in_m} \gls{in_m}},
	sort={zz m c arr},
}

\newglossaryentry{trip_rate_out_t_c}
{
	type=amod,
	name={\ensuremath{\gls{trip_rate_m}^{\gls{t},\gls{c},\gls{out}}}},
	description={number of vehicles with \glsdesc{c} \gls{c} arriving at \glsdesc{t} \gls{t} after serving  \glsdesc{in_m} \gls{in_m}},
	sort={zz m c dep},
}

\newglossaryentry{dist_vehicle}
{
	name={\ensuremath{N}},
	description={vehicle distribution}
}

\newglossaryentry{dist_vehicle_initial}
{
	name={\ensuremath{\gls{dist_vehicle}_I}},
	sort={N I},
	description={initial \glsdesc{dist_vehicle}}
}

\newglossaryentry{dist_vehicle_final}
{
	name={\ensuremath{\gls{dist_vehicle}_F}},
	sort={N F},
	description={final \glsdesc{dist_vehicle}}
}

\newglossaryentry{map_fun}
{
	name={\ensuremath{\mathcal{M}}},
	description={xx}
}

\newglossaryentry{map_fun_s_con}
{
	type=amod,
	name={\ensuremath{\gls{map_fun}_{\gls{charger_set},\gls{con_load_set}}}},
	sort={M S L},
    	description={maps a charging station
$\gls{s}$ to the associated controllable load $\gls{in_con_load}$ and distribution
network $\gls{in_dist}$}
}

\newglossaryentry{in_bus_map_fun_s_con}
{
	name={\ensuremath{\gls{in_bus}_{\gls{map_fun_s_con}(\gls{s})}}},
	sort={n M S L},
	description={\glsdesc{in_bus} in \glsdesc{in_dist} to which \glsdesc{in_charger} $\gls{in_charger} \in \gls{charger_set}$ is connected}
}

\newglossaryentry{in_dist_map_fun_s_con}
{
	name={\ensuremath{\gls{in_dist}_{\gls{map_fun_s_con}(\gls{s})}}},
	sort={d M S L},
	description={\glsdesc{in_dist} to which \glsdesc{in_charger} $\gls{in_charger} \in \gls{charger_set}$ is connected}
}

\newglossaryentry{msg}
{
	type=amod,
	name={\ensuremath{\gls{map_fun}_{\gls{charger_set},\gls{exp_edge_set_charger}}}},
	sort={M S As},
	description={maps a charging
	station $\gls{s}$ for each time step $\gls{t}$ to all arcs in $\gls{exp_edge_set_charger}$ that
	represent charging vehicles at this station}
}

\newglossaryentry{bigO}
{
	name={\ensuremath{\mathcal{O}}},
	description={xx}
}

{
	\newglossaryentry{E_total}
	{
		name={\ensuremath{\gls{E}_{\mathrm{total}}}},
		sort={E total},
		description={total energy consumption}
	}

	\newglossaryentry{E_total_base}
	{
		name={\ensuremath{\gls{E}_{\mathrm{total,base}}}},
		sort={E total},
		description={total energy consumption in the base case}
	}

	\newglossaryentry{E_amod_charge}
	{
		name={\ensuremath{\gls{E}_{\gls{charge},\mathrm{\gls{amod}}}}},
		sort={E charge  AMoD},
		description={energy delivered to the charging stations}
	}

	\newglossaryentry{E_amod}
	{
		name={\ensuremath{\gls{E}_{\mathrm{\gls{amod}}}}},
		sort={E AMoD},
		description={additional energy consumption caused by the \gls{eamod} system}
	}

	\newglossaryentry{E_amod_losses}
	{
		name={\ensuremath{\gls{E}_{\gls{loss},\mathrm{\gls{amod}}}}},
		sort={E loss AMoD},
		description={link losses caused by the \gls{eamod} system}
	}

	\newglossaryentry{electricity_cost_amod}
	{
		name={\ensuremath{\gls{price}_{\gls{el},\mathrm{\gls{amod}}}}},
		sort={V el amod},
		description={Electricity cost, \gls{amod}}
	}

	\newglossaryentry{electricity_cost_charging}
	{
		name={\ensuremath{\gls{price}_{\gls{el},\mathrm{charge,\gls{amod}}}}},
		sort={V el charge amod},
		description={Electricity cost, charging}
	}

	\newglossaryentry{electricity_cost_losses}
	{
		name={\ensuremath{\gls{price}_{\gls{el},\mathrm{loss ,\gls{amod}}}}},
		sort={V el losses amod},
		description={Electricity cost, losses}
	}

}

{
\newglossaryentry{reals}
{
	name={\ensuremath{\mathbb{R}}},
	description={real numbers}
}

\newglossaryentry{reals_plus}
{
	name={\ensuremath{\gls{reals}^{+}}},
	description={non-negative real numbers}
}

\newglossaryentry{complexes}
{
	name={\ensuremath{\mathbb{C}}},
	description={complex numbers}
}

\newglossaryentry{zeros}
{
	name={\ensuremath{\mathbb{0}}},
	description={matrix of all zeros}
}

\newglossaryentry{ones}
{
	name={\ensuremath{\mathbb{1}}},
	description={matrix of all ones}
}

\newglossaryentry{eye}
{
	name={\ensuremath{\mat{I}}},
	description={identity matrix}
}

\newglossaryentry{hermitians}
{
	name={\ensuremath{\mathbb{H}}},
	description={hermitian matrices}
}

\newglossaryentry{hermitians_psd}
{
	name={\ensuremath{\mathbb{H}_+}},
	description={set of positive semi-definite hermitian matrices}
}

\newglossaryentry{j}
{
	name={\ensuremath{j}},
	description={imaginary unit}
}

\newglossaryentry{indicator}
{
	name={\ensuremath{\mathbb{1}}},
	description={indicator function}
}

\newglossaryentry{leq_lin}
{
	name={\ensuremath{\leq_{\gls{lin}}}},
	description={linearized constraint}
}

\newglossaryentry{foft}
{
	name={\ensuremath{\lbrack\gls{t}\rbrack}},
	description={xx}
}

}

{
\newglossaryentry{bus_set}
{
	type=opf,
	name={\ensuremath{\mathcal{N}}},
	sort={N},
	description={set of buses}
}

\newglossaryentry{bus_set_plus}
{
	name={\ensuremath{\gls{bus_set}^{+}}},
	sort={N},
	description={\glsdesc{bus_set} excluding the reference bus}
}

\newglossaryentry{bus_set_dist_plus}
{
	name={\ensuremath{\gls{bus_set_plus}_{\gls{in_dist}}}},
	description={xx}
}

\newglossaryentry{n_bus}
{
	name={\ensuremath{N}},
	sort={N},
	description={number of buses}
}

\newglossaryentry{n_bus_3}
{
	name={\ensuremath{N_{3\gls{in_phase}}}},
	description={number of three-phase buses}
}

\newglossaryentry{n_bus_phase}
{
	name={\ensuremath{N_{\gls{phase_set}}}},
	description={total number of phases in all buses}
}

\newglossaryentry{in_bus}
{
	name={\ensuremath{n}},
	description={bus}
}

\newglossaryentry{in_bus_pre}
{
	name={\ensuremath{m}},
	description={xx}
}

\newglossaryentry{in_bus_post}
{
	name={\ensuremath{o}},
	description={xx}
}

\newglossaryentry{in_line}
{
	name={\ensuremath{\gls{in_bus},\gls{in_bus_post}}},
	description={link}
}

\newglossaryentry{in_line_pre}
{
	name={\ensuremath{\gls{in_bus_pre},\gls{in_bus}}},
	description={xx}
}

\newglossaryentry{line_set}
{
	type=opf,
	name={\ensuremath{\mathcal{E}}},
	sort={E},
	description={set of links}
}
}

{
\newglossaryentry{v}
{
	name={\ensuremath{v}},
	description={complex voltage}
}

\newglossaryentry{u}
{
	name={\ensuremath{u}},
	description={voltage magnitude}
}

\newglossaryentry{u_vec}
{
	name={\ensuremath{\vec{u}}},
	description={xx}
}

\newglossaryentry{theta}
{
	name={\ensuremath{\theta}},
	description={voltage angle}
}
\newglossaryentry{theta_vec}
{
	name={\ensuremath{\vec{\theta}}},
	description={xx}
}

\newglossaryentry{s}
{
	name={\ensuremath{s}},
	description={complex power}
}

\newglossaryentry{s_vec}
{
	name={\ensuremath{\vec{s}}},
	description={complex power}
}

\newglossaryentry{p}
{
	name={\ensuremath{p}},
	description={active power}
}

\newglossaryentry{p_vec}
{
	name={\ensuremath{\vec{p}}},
	description={xx}
}

\newglossaryentry{q}
{
	name={\ensuremath{q}},
	description={reactive power}
}

\newglossaryentry{q_vec}
{
	name={\ensuremath{\vec{q}}},
	description={xx}
}

\newglossaryentry{E}
{
	name={\ensuremath{E}},
	description={energy}
}

\newglossaryentry{y_bus}
{
	name={\ensuremath{\mat{Y}}},
	description={network bus admittance matrix}
}

\newglossaryentry{z_mat}
{
	name={\ensuremath{\mat{Z}}},
	description={impedance matrix}
}

\newglossaryentry{z_mat_line}
{
	type=opf,
	name={\ensuremath{\gls{z_mat}_{\gls{in_line}}}},
	sort={Z n o},
	description={\glsdesc{z_mat} of link $(\gls{in_line})\in\gls{line_set}$}
}

\newglossaryentry{y_mat}
{
	name={\ensuremath{\mat{Y}}},
	description={admittance matrix}
}

\newglossaryentry{power_factor}
{
	name={\ensuremath{\eta}},
	description={power factor}
}

\newglossaryentry{power_factor_bar}
{
	name={\ensuremath{\tilde{\eta}}},
	description={power factor}
}

\newglossaryentry{power_factor_vec}
{
	name={\ensuremath{\vec{\eta}}},
	description={power factor}
}

}

{
\newglossaryentry{n_dist_net}
{
	name={\ensuremath{D}},
	sort={D},
	description={number of \glsentrydescplural{dist_net}}
}

\newglossaryentry{dist_net}
{
	name={\ensuremath{P}},
	description={radial distribution network},
    descriptionplural={radial distribution networks}
}

\newglossaryentry{dist_net_d}
{	
	name={\ensuremath{\gls{dist_net}_{\gls{in_dist}}}},
	description={distribution network}
}

\newglossaryentry{in_phase}
{
	name={\ensuremath{\phi}},
	description={phase}
}

\newglossaryentry{phase_set}
{
	type=opf,
	name={\ensuremath{\Phi}},
	sort={zz P},
	description={set of phases}
}

\newglossaryentry{phase_set_bus}
{
	name={\ensuremath{\gls{phase_set}_{\gls{in_bus}}}},
	sort={zz P n},
	description={\glsdesc{phase_set} in bus $\gls{in_bus} \in \gls{bus_set}$}
}

\newglossaryentry{phase_set_bus_d}
{
	name={\ensuremath{\gls{phase_set}_{\gls{in_bus},\gls{in_dist}}}},
	description={xx}
}

\newglossaryentry{n_phase_set}
{
	name={\ensuremath{|\gls{phase_set}|}},
	description={number of phases in bus \gls{in_bus}}
}

\newglossaryentry{n_phase_set_bus}
{
	name={\ensuremath{|\gls{phase_set_bus}|}},
	description={number of phases in bus \gls{in_bus}}
}

\newglossaryentry{phase_set_delta_bus}
{
	name={\ensuremath{\gls{phase_set_bus}^\Delta}},
	description={xx}
}

\newglossaryentry{n_phase_set_delta_bus}
{
	name={\ensuremath{|\gls{phase_set_delta_bus}|}},
	description={xx}
}

\newglossaryentry{n_phase_set_line}
{
	name={\ensuremath{|\gls{phase_set_line}|}},
	description={number of phases in bus \gls{in_bus}}
}

\newglossaryentry{phase_set_line}
{
	name={\ensuremath{\gls{phase_set}_{\gls{in_line}}}},
	sort={zz P n o},
	description={\glsdesc{phase_set} in link $(\gls{in_line}) \in \gls{line_set}$}
}

\newglossaryentry{u_bus_phase}
{
	name={\ensuremath{\gls{u}_{\gls{in_bus}}^{\gls{in_phase}}}},
	description={voltage magnitude at phase \gls{in_phase} in bus \gls{in_bus}}
}

\newglossaryentry{u_bus_phase_d}
{
	name={\ensuremath{\gls{u}_{\gls{in_bus},\gls{in_dist}}^{\gls{in_phase}}}},
	description={xx}
}

\newglossaryentry{v_bus_phase}
{
	type=opf,
	name={\ensuremath{\gls{v}_{\gls{in_bus}}^{\gls{in_phase}}}},
	sort={v n o},
	description={\glsdesc{v} at phase $\gls{in_phase}\in\gls{phase_set_bus}$ in bus $\gls{in_bus}\in\gls{bus_set}$}
}

\newglossaryentry{v_vec}
{
    name={\ensuremath{\vec{v}}},
	description={xx}
}

\newglossaryentry{v_bus}
{
	name={\ensuremath{\gls{v_vec}_n}},
	description={complex voltage at bus \gls{in_bus}}
}

\newglossaryentry{v_dist_net}
{
	name={\ensuremath{^d\gls{v_vec}}},
	description={xx}
}

\newglossaryentry{s_inj}
{
	name={\ensuremath{\gls{s}_{\gls{inj}}}},
	description={complex power injection}
}

\newglossaryentry{s_inj_vec}
{
	name={\ensuremath{\gls{s_vec}_{\gls{inj}}}},
	description={xx}
}

\newglossaryentry{s_inj_bus_phase}
{
	type=opf,
	name={\ensuremath{\gls{s}_{\gls{inj},n}^{\gls{in_phase}}}},
	sort={s inj n},
	description={\glsdesc{s_inj} at phase $\gls{in_phase}\in\gls{phase_set_bus}$ in bus $\gls{in_bus}\in\gls{bus_set}$}
}

\newglossaryentry{s_inj_bus}
{
	name={\ensuremath{\vec{s}_{\gls{inj},n}}},
	description={power injection at bus \gls{in_bus}}
}

\newglossaryentry{s_inj_bus_pq}
{
	name={\ensuremath{\vec{s}_{\gls{inj},\gls{p_load},n}}},
	description={xx}
}

\newglossaryentry{p_inj_vec}
{
	name={\ensuremath{\gls{p_vec}_{\gls{inj}}}},
	description={xx}
}
\newglossaryentry{q_inj_vec}
{
	name={\ensuremath{\gls{q_vec}_{\gls{inj}}}},
	description={xx}
}

\newglossaryentry{s_inj_dist_net}
{
	name={\ensuremath{^d\vec{s}_{\gls{inj}} }},
	description={xx}
}

}

{
\newglossaryentry{obj_fun}
{
	name={\ensuremath{f}},
	description={objective function}
}

\newglossaryentry{unc_load_s}
{
	name={\ensuremath{\vec{s}_{\gls{unc}}}},
	description={complex power of uncontrollable load}
}

\newglossaryentry{unc_load_s_bus}
{
	type=opf,
	name={\ensuremath{\vec{s}_{\gls{unc},\gls{in_bus}}}},
	sort={s unc n},
	description={\glsdesc{unc_load_s} in bus $\gls{in_bus}\in \gls{bus_set_plus}$}
}

\newglossaryentry{unc_load_s_p_y_bus}
{
	name={\ensuremath{\vec{s}_{\gls{unc},\gls{p_load},\gls{y_con},n}}},
	description={\glsdesc{y_con} constant power load}
}

\newglossaryentry{unc_load_s_z_bus}
{
	name={\ensuremath{\vec{s}_{\gls{unc},\gls{z_load},n}}},
	description={constant impedance load}
}

\newglossaryentry{unc_load_y_z_y}
{
	name={\ensuremath{y_{\gls{unc},\gls{y_con},n}}},
	description={xx}
}
\newglossaryentry{unc_load_y_z_y_bus}
{
	name={\ensuremath{\vec{y}_{\gls{unc},\gls{y_con},n}}},
	description={admittance of the \glsdesc{y_con} \glsdesc{unc_load_s_z_bus}}
}

\newglossaryentry{unc_load_y_z_d}
{
	name={\ensuremath{y_{\gls{unc},\gls{d_con},n}}},
	description={admittance of the \glsdesc{d_con} \glsdesc{unc_load_s_z_bus}}
}

\newglossaryentry{unc_load_y_z_d_bus}
{
	name={\ensuremath{\vec{y}_{\gls{unc},\gls{d_con},n}}},
	description={admittance of the \glsdesc{d_con} \glsdesc{unc_load_s_z_bus}}
}

\newglossaryentry{unc_load_y_mat_z}
{
	name={\ensuremath{\gls{y_mat}_{\gls{unc},n}}},
	description={xx}
}

\newglossaryentry{unc_load_y_mat_z_y}
{
	name={\ensuremath{\gls{y_mat}_{\gls{unc},\gls{y_con},n}}},
	description={\glsdesc{y_mat} for the \glsdesc{y_con} \glsdesc{unc_load_s_z_bus}}
}
\newglossaryentry{unc_load_y_mat_z_d}
{
	name={\ensuremath{\gls{y_mat}_{\gls{unc},\gls{d_con},n}}},
	description={\glsdesc{y_mat} for the \glsdesc{d_con} \glsdesc{unc_load_s_z_bus}}
}

\newglossaryentry{in_con_load}
{
	name={\ensuremath{\ell}},
	description={controllable load}
}

\newglossaryentry{con_load_set}
{
	type=opf,
	name={\ensuremath{\mathcal{L}}},
	sort={L},
	description={set of controllable loads}
}

\newglossaryentry{con_load_set_d}
{
	name={\ensuremath{\mathcal{L}_{\gls{in_dist}}}},
	description={xx}
}

\newglossaryentry{n_con_load}
{
	name={\ensuremath{L}},
	sort={L},
	description={number of controllable loads}
}

\newglossaryentry{con_load_s}
{
	name={\ensuremath{\gls{s}_{\gls{con}}}},
	description={complex power of controllable load}
}

\newglossaryentry{con_load_s_vec}
{
	name={\ensuremath{\vec{s}_{\gls{con}}}},
	description={complex power of controllable load}
}

\newglossaryentry{con_load_s_c}
{	
	name={\ensuremath{\gls{s}_{\gls{con},\gls{in_con_load}}}},
	sort={s con l},
	description={\glsdesc{s} of controllable load $\gls{in_con_load} \in \gls{con_load_set}$}
}

\newglossaryentry{con_load_s_c_set}
{
	name={\ensuremath{\gls{s_set}_{\gls{con},\gls{in_con_load}}}},
	description={xx}
}

\newglossaryentry{con_load_s_c_vec}
{
	type=opf,
	name={\ensuremath{\vec{s}_{\gls{con},\gls{in_con_load}}}},
	sort={s con l},
	description={\glsdesc{s} of controllable load $\gls{in_con_load} \in \gls{con_load_set}$}
}

\newglossaryentry{con_load_s_c_phase}
{
	name={\ensuremath{\gls{con_load_s_c}^{\gls{in_phase}}}},
	description={xx}
}

\newglossaryentry{con_load_bus}
{
	type=opf,
	name={\ensuremath{n_{\gls{in_con_load}}}},
	sort={n l},
	description={reference bus of controllable load $\gls{in_con_load} \in \gls{con_load_set}$}
}

\newglossaryentry{phase_set_con_load}
{
	name={\ensuremath{\gls{phase_set}_{\gls{con_load_bus}}}},
	description={xx}
}

\newglossaryentry{con_load_p}
{
	name={\ensuremath{\gls{p}_{\gls{con}}}},
	description={xx}
}

\newglossaryentry{con_load_q}
{
	name={\ensuremath{\gls{q}_{\gls{con}}}},
	description={xx}
}

\newglossaryentry{con_load_p_bus}
{
	name={\ensuremath{\gls{p_vec}_{\gls{con},\gls{in_con_load}}}},
	description={xx}
}

\newglossaryentry{con_load_p_bus_phase}
{
	name={\ensuremath{\gls{p}_{\gls{con},\gls{in_con_load}}^{\gls{in_phase}}}},
	description={xx}
}

\newglossaryentry{con_load_p_bus_phase_d}
{
	name={\ensuremath{\gls{p}_{\gls{con},\gls{in_con_load},\gls{in_dist}}^{\gls{in_phase}}}},
	description={xx}
}

\newglossaryentry{con_load_p_bus_phase_min}
{
	name={\ensuremath{\gls{p}_{\gls{con},\gls{min},\gls{in_con_load}}^{\gls{in_phase}}}},
	description={xx}
}

\newglossaryentry{con_load_p_bus_phase_max}
{
	name={\ensuremath{\gls{p}_{\gls{con},\gls{max},\gls{in_con_load}}^{\gls{in_phase}}}},
	description={xx}
}

\newglossaryentry{con_load_q_bus}
{
	name={\ensuremath{\gls{q_vec}_{\gls{con},\gls{in_con_load}}}},
	description={xx}
}

\newglossaryentry{con_load_q_bus_phase}
{
	name={\ensuremath{\gls{q}_{\gls{con},\gls{in_con_load}}^{\gls{in_phase}}}},
	description={xx}
}

\newglossaryentry{con_load_q_bus_phase_min}
{
	name={\ensuremath{\gls{q}_{\gls{con},\gls{min},\gls{in_con_load}}^{\gls{in_phase}}}},
	description={xx}
}

\newglossaryentry{con_load_q_bus_phase_max}
{
	name={\ensuremath{\gls{q}_{\gls{con},\gls{max},\gls{in_con_load}}^{\gls{in_phase}}}},
	description={xx}
}

\newglossaryentry{y_mat_line}
{
	name={\ensuremath{\gls{y_mat}_{\gls{in_line}}}},
	description={phase admittance matrix of link (\gls{in_line})}
}

\newglossaryentry{y_mat_shunt}
{
	name={\ensuremath{\gls{y_mat}}},
	description={shunt admittance matrix}
}

\newglossaryentry{y_mat_shunt_bus}
{
	type=opf,
	name={\ensuremath{\gls{y_mat_shunt}_{\gls{in_bus}}}},
	sort={y n},
	description={\glsdesc{y_mat_shunt} of bus $\gls{in_bus}\in\gls{bus_set}$}
}

}

{
\newglossaryentry{v_mag_min}
{
    name={\ensuremath{\gls{u}_{\gls{min}}}},
    description={xx}
}

\newglossaryentry{v_mag_max}
{
    name={\ensuremath{\gls{u}_{\gls{max}}}},
    description={xx}
}

\newglossaryentry{v_mag_bus_phase_min}
{
    name={\ensuremath{\gls{u}^{\gls{in_phase}}_{\gls{min},\gls{in_bus}}}},
    description={xx}
}

\newglossaryentry{v_mag_bus_phase_min_d}
{
    name={\ensuremath{\gls{u}^{\gls{in_phase}}_{\gls{min},\gls{in_bus},\gls{in_dist}}}},
    description={xx}
}

\newglossaryentry{v_mag_bus_phase_max}
{
    name={\ensuremath{\gls{u}^{\gls{in_phase}}_{\gls{max},\gls{in_bus}}}},
    description={xx}
}

\newglossaryentry{v_mag_bus_phase_max_d}
{
    name={\ensuremath{\gls{u}^{\gls{in_phase}}_{\gls{max},\gls{in_bus},\gls{in_dist}}}},
    description={xx}
}

\newglossaryentry{u_viol}
{
    sort={u viol},
    name={\ensuremath{\gls{u}_{\gls{viol}, \gls{int}}}},
    description={integral absolute voltage magnitude constraint violation}
}

\newglossaryentry{u_viol_bus_phase_d}
{
    sort={u viol x},
    name={\ensuremath{\gls{u}^{\gls{in_phase}}_{\gls{viol},\gls{in_bus},\gls{in_dist}}}},
    description={voltage magnitude constraint violation at phase $\gls{in_phase}\in\gls{phase_set_bus}$ in bus $\gls{in_bus}\in\gls{bus_set}$}
}

{
\newglossaryentry{con_load_s_c_min}
{
    name={\ensuremath{\gls{con_load_s}{}_{,\gls{min},\gls{in_con_load}}}},
    description={xx}
}

\newglossaryentry{con_load_s_c_max}
{
    name={\ensuremath{\gls{con_load_s}{}_{,\gls{max},\gls{in_con_load}}}},
    description={xx}
}

\newglossaryentry{con_load_power_factor_phase}
{
    name={\ensuremath{\gls{power_factor}_{\gls{in_con_load}}^{\gls{in_phase}}}},
    description={xx}
}

\newglossaryentry{con_load_power_factor_min_phase}
{
    name={\ensuremath{\gls{power_factor}_{\gls{min},\gls{in_con_load}}^{\gls{in_phase}}}},
    description={xx}
}

\newglossaryentry{con_load_power_factor_bar_min_phase}
{
    name={\ensuremath{\gls{power_factor_bar}_{\gls{min},\gls{in_con_load}}^{\gls{in_phase}}}},
    description={xx}
}

\newglossaryentry{con_load_power_factor_est_phase}
{
    name={\ensuremath{\gls{power_factor}_{\textrm{est},\gls{in_con_load}}^{\gls{in_phase}}}},
    description={estimated power factor of the controllable load}
}

\newglossaryentry{con_load_s_c_phase_mag_max}
{
    name={\ensuremath{\gls{s}_{\gls{con},\gls{mag},\gls{max},\gls{in_con_load}}^{\gls{in_phase}}}},
    description={xx}
}

\newglossaryentry{con_load_p_phase_delta_min}
{
    name={\ensuremath{\Delta\gls{con_load_p}^{\gls{in_phase}}{}_{,\gls{min},\gls{in_con_load}}}},
    description={xx}
}

\newglossaryentry{con_load_p_phase_delta_max}
{
    name={\ensuremath{\Delta\gls{con_load_p}^{\gls{in_phase}}{}_{,\gls{max},\gls{in_con_load}}}},
    description={xx}
}

\newglossaryentry{con_load_q_phase_delta_min}
{
    name={\ensuremath{\Delta\gls{con_load_q}^{\gls{in_phase}}{}_{,\gls{min},\gls{in_con_load}}}},
    description={xx}
}

\newglossaryentry{con_load_q_phase_delta_max}
{
    name={\ensuremath{\Delta\gls{con_load_q}^{\gls{in_phase}}{}_{,\gls{max},\gls{in_con_load}}}},
    description={xx}
}

}

{

\newglossaryentry{v_pcc_vec}
{
	name={\ensuremath{\gls{v_vec}_{\gls{pcc}}}},
	description={xx}
}

\newglossaryentry{v_pcc_phase}
{
	name={\ensuremath{\gls{v}_{\gls{pcc}}^{\gls{in_phase}}}},
	description={xx}
}

\newglossaryentry{v_ref_phase}
{
	name={\ensuremath{\gls{v}_{\gls{ref}}^{\gls{in_phase}}}},
	sort={v ref},
	description={reference \glsdesc{v} for the network at \glsdesc{in_phase} $\gls{in_phase}\in\gls{phase_set}$}
}

\newglossaryentry{v_ref_vec}
{
	name={\ensuremath{\gls{v_vec}_{\gls{ref}}}},
	description={xx}
}

\newglossaryentry{s_pcc}
{
	name={\ensuremath{\gls{s}_{\gls{pcc}}}},
	description={complex power supplied by the utility at the substation}
}

\newglossaryentry{s_pcc_d}
{
	name={\ensuremath{\gls{s}_{\gls{pcc},\gls{in_dist}}}},
	description={xx}
}

\newglossaryentry{p_pcc}
{
	name={\ensuremath{\gls{p}_{\gls{pcc}}}},
	description={real power supplied by the utility at the substation}
}

\newglossaryentry{p_pcc_d}
{
	name={\ensuremath{\gls{p}_{\gls{pcc},\gls{in_dist}}}},
	description={xx}
}

\newglossaryentry{s_pcc_vec}
{
	name={\ensuremath{\vec{s}_{\gls{pcc}}}},
	description={xx}
}

\newglossaryentry{s_set}
{
	name={\ensuremath{\mathcal{S}}},
	description={xx}
}

\newglossaryentry{s_pcc_set}
{
	name={\ensuremath{\gls{s_set}_{\gls{pcc}}}},
	description={xx}
}

\newglossaryentry{s_pcc_vec_min}
{
	name={\ensuremath{\gls{s_pcc_vec}{}_{,\gls{min}}}},
	description={xx}
}

\newglossaryentry{s_pcc_vec_max}
{
	name={\ensuremath{\gls{s_pcc_vec}{}_{,\gls{max}}}},
	description={xx}
}

\newglossaryentry{s_pcc_phase}
{
	name={\ensuremath{\gls{s}_{\gls{pcc}}^{\gls{in_phase}}}},
	description={xx}
}

\newglossaryentry{p_pcc_phase}
{
	name={\ensuremath{\gls{p}_{\gls{pcc}}^{\gls{in_phase}}}},
	description={xx}
}

\newglossaryentry{q_pcc_phase}
{
	name={\ensuremath{\gls{q}_{\gls{pcc}}^{\gls{in_phase}}}},
	description={xx}
}

\newglossaryentry{p_pcc_phase_min}
{
	name={\ensuremath{\gls{p}_{\gls{pcc},\gls{min}}^{\gls{in_phase}}}},
	description={xx}
}

\newglossaryentry{p_pcc_phase_max}
{
	name={\ensuremath{\gls{p}_{\gls{pcc},\gls{max}}^{\gls{in_phase}}}},
	description={xx}
}

\newglossaryentry{q_pcc_phase_min}
{
	name={\ensuremath{\gls{q}_{\gls{pcc},\gls{min}}^{\gls{in_phase}}}},
	description={xx}
}

\newglossaryentry{q_pcc_phase_max}
{
	name={\ensuremath{\gls{q}_{\gls{pcc},\gls{max}}^{\gls{in_phase}}}},
	description={xx}
}

\newglossaryentry{power_factor_pcc_min_phase}
{
    name={\ensuremath{\gls{power_factor}_{\gls{pcc},\gls{min}}^{\gls{in_phase}}}},
    description={xx}
}

\newglossaryentry{power_factor_bar_pcc_min_phase}
{
    name={\ensuremath{\gls{power_factor_bar}_{\gls{pcc},\gls{min}}^{\gls{in_phase}}}},
    description={xx}
}

\newglossaryentry{power_factor_3_pcc_est}
{
    name={\ensuremath{\xi_{\gls{pcc},\textrm{est}}}},
    description={xx}
}

\newglossaryentry{power_factor_3_pcc}
{
    name={\ensuremath{\xi_{\gls{pcc}}}},
    description={xx}
}

}

{
\newglossaryentry{v_sq_mat}
{
	name={\ensuremath{\vec{W}}},
	description={xx}
}
\newglossaryentry{s_mat}
{
	name={\ensuremath{\vec{S}}},
	description={xx}
}

\newglossaryentry{ell_mat}
{
	name={\ensuremath{\vec{L}}},
	description={xx}
}

\newglossaryentry{labmda_vec}
{
	name={\ensuremath{\vec{\Lambda}}},
	description={vector of power flows}
}

\newglossaryentry{labmda_phase}
{
	name={\ensuremath{\Lambda^{\gls{in_phase}}}},
	description={xx}
}

\newglossaryentry{i}
{
	name={\ensuremath{i}},
	description={complex current}
}

\newglossaryentry{i_vec}
{
	name={\ensuremath{\vec{i}}},
	description={xx}
}

\newglossaryentry{i_line}
{
	name={\ensuremath{\gls{i_vec}_{\gls{in_line}}}},
	description={\glsdesc{i} through link (\gls{in_line})}
}

\newglossaryentry{i_line_phase}
{
	type=opf,
	name={\ensuremath{\gls{i}_{\gls{in_line}}^{\gls{in_phase}}}},
	sort={i n o},
	description={\glsdesc{i} through link $(\gls{in_line})\in\gls{line_set}$}
}

\newglossaryentry{i_line_mag_max}
{
	name={\ensuremath{\gls{i}_{\gls{mag},\gls{max},\gls{in_line}}}},
	description={xx}
}

\newglossaryentry{i_line_phase_mag_max}
{
	name={\ensuremath{\gls{i}_{\gls{mag},\gls{max},\gls{in_line}}^{\gls{in_phase}}}},
	description={xx}
}

\newglossaryentry{alpha}
{
	name={\ensuremath{\alpha}},
	description={xx}
}

\newglossaryentry{delta_mat}
{
	name={\ensuremath{\vec{\Delta}}},
	description={xx}
}

\newglossaryentry{gamma_mat}
{
	name={\ensuremath{\mat{\Gamma}}},
	description={matrix of fixed voltage ratios}
}

\newglossaryentry{gamma_mat_line}
{
	name={\ensuremath{\gls{gamma_mat}_{\gls{in_line}}}},
	sort={zz G n o},
	description={matrix of fixed voltage ratios for \glsdesc{in_line} $(\gls{in_line})\in\gls{line_set}$}
}

\newglossaryentry{u_bus_phase_fix}
{
	name={\ensuremath{\tilde{\gls{u}}_{\gls{in_bus}}^{\gls{in_phase}}}},
	description={approximate voltage magnitude}
}

\newglossaryentry{v_bus_fix}
{
	name={\ensuremath{\tilde{\gls{v_vec}}_{\gls{in_bus}}}},
	sort={v n o},
	description={voltage used to determine the fixed voltage ratios in \glsdesc{in_bus} $\gls{in_bus}\in\gls{bus_set_plus}$}
}

\newglossaryentry{i_line_fix}
{
	name={\ensuremath{\tilde{\gls{i_vec}}_{\gls{in_line}}}},
	sort={i n o},
	description={fixed link current in link $(\gls{in_line})\in\gls{line_set}$ used to determine the fixed link losses}
}

\newglossaryentry{ell_mat_fix}
{
	name={\ensuremath{\tilde{\gls{ell_mat}}}},
	description={fixed \glsdesc{ell_mat}}
}

}

{
	\newglossaryentry{network_state}
	{
		name={\ensuremath{\vec{x}}},
		description={network state}
	}

	\newglossaryentry{A_mat}
	{
		name={\ensuremath{\mat{A}}},
		description={xx}
	}

	\newglossaryentry{pf_manifold}
	{
		name={\ensuremath{\mathcal{M}}},
		description={power flow manifold}
	}

}

{
	\newglossaryentry{n_poly}
	{
		name={\ensuremath{n_{\textrm{poly}}}},
		description={number of faces in polygon}
	}

	\newglossaryentry{A_poly}
	{
		name={\ensuremath{\mat{A}_{\textrm{poly},\gls{n_poly}}}},
		description={xx}
	}

	\newglossaryentry{b_poly}
	{
		name={\ensuremath{\vec{b}_{\textrm{poly},\gls{n_poly}}}},
		description={xx}
	}

	\newglossaryentry{theta_poly}
	{
		name={\ensuremath{\theta_{\textrm{poly}}}},
		description={xx}
	}
	
}

{
	\newglossaryentry{s_pcc_sum_mag_max}
	{
		name={\ensuremath{\hat{\gls{s}}_{\gls{pcc}}}},
		sort={s 0},
		description={substation transformer rating}
	}

	\newglossaryentry{s_pcc_sum_mag_max_d}
	{
		name={\ensuremath{\hat{\gls{s}}_{\gls{pcc},\gls{in_dist}}}},
		description={xx}
	}

	\newglossaryentry{s_pcc_sum_mag_max_viol}
	{
		name={\ensuremath{\hat{\gls{s}}_{\gls{pcc},\gls{viol}, \gls{int}}}},
		sort={s 0 viol},
		description={integral substation transformer rating violation}
	}

	\newglossaryentry{s_pcc_sum_mag_max_viol_d}
	{
		name={\ensuremath{\hat{\gls{s}}_{\gls{pcc},\gls{in_dist},\gls{viol}}}},
		sort={s 0 viol},
		description={substation transformer rating violation for $\gls{in_dist} \in \gls{in_dist_set}$}
	}
}

{
	\newglossaryentry{in_dist}
	{
		name={\ensuremath{d}},
		description={distribution network},
		descriptionplural={distribution networks},
	}

	\newglossaryentry{n_dist}
	{
		name={\ensuremath{D}},
		sort={D},
		description={number of \glsdescplural{in_dist}}
	}

	\newglossaryentry{in_dist_set}
	{
		type=opf,
		name={\ensuremath{\mathcal{\gls{n_dist}}}},
		sort={D},
		description={set of \glsdescplural{in_dist}}
	}

	\newglossaryentry{in_dist_sup}
	{
		name={\ensuremath{{}^{\gls{in_dist}}}},
		description={xx}
	}

}

\newglossaryentry{p_load}
{
	name={\ensuremath{\mathrm{PQ}}},
	description={constant power load part}
}

\newglossaryentry{z_load}
{
	name={\ensuremath{\mathrm{Z}}},
	description={constant impedance part}
}

\newglossaryentry{y_con}
{
	name={\ensuremath{\mathrm{y}}},
	description={wye-connected}
}

\newglossaryentry{d_con}
{
	name={\ensuremath{\mathrm{\Delta}}},
	description={delta-connected}
}

\newglossaryentry{inj}
{
	name={\ensuremath{\mathrm{inj}}},
	description={injection}
}

\newglossaryentry{con}
{
	name={\ensuremath{\mathrm{con}}},
	description={controllable}
}

\newglossaryentry{unc}
{
	name={\ensuremath{\mathrm{unc}}},
	description={uncontrollable}
}

\newglossaryentry{min}
{
	name={\ensuremath{\mathrm{min}}},
	description={xx}
}

\newglossaryentry{max}
{
	name={\ensuremath{\mathrm{max}}},
	description={xx}
}

\newglossaryentry{mag}
{
	name={\ensuremath{\mathrm{mag}}},
	description={magnitude}
}

\newglossaryentry{sum}
{
	name={\ensuremath{\mathrm{sum}}},
	description={xx}
}

\newglossaryentry{lin}
{
	name={\ensuremath{\mathrm{lin}}},
	description={xx}
}

\newglossaryentry{base}
{
	name={\ensuremath{\mathrm{base}}},
	description={xx}
}

\newglossaryentry{loss}
{
	name={\ensuremath{\mathrm{loss}}},
	description={xx}
}

\newglossaryentry{charge}
{
	name={\ensuremath{\mathrm{charge}}},
	description={xx}
}

\newglossaryentry{pcc}
{
	name={\ensuremath{\mathrm{0}}},
	description={relative to reference bus}
}

\newglossaryentry{ref}
{
	name={\ensuremath{\mathrm{ref}}},
	description={reference}
}

\newglossaryentry{el}
{	
	name={\ensuremath{\mathrm{el}}},
	description={electricity}
}

\newglossaryentry{viol}
{	
	name={\ensuremath{\mathrm{viol}}},
	description={violation}
}

\newglossaryentry{int}
{	
	name={\ensuremath{\mathrm{int}}},
	description={integral}
}

\newglossaryentry{out}
{
	name={\ensuremath{\mathrm{arr}}},
	description={xx}
}

\newglossaryentry{in}
{
	name={\ensuremath{\mathrm{dep}}},
	description={xx}
}
\newglossaryentry{gld}
{
		name={GridLAB-D},
	description={xx}
}

\newglossaryentry{eamod}
{
	name={electric \gls{amod}},
	description={electric \gls{amod}}
}

\newglossaryentry{amod_opf}
{
		name={\acrshort{amod}-\acrshort{opf}},
	description={xx}
}

\newglossaryentry{mopf}
{
		name={multi-\gls{opf}},
	description={xx}
}

\newglossaryentry{eq:powerflow}
{
		name={\text{power flow equation}},
	description={xx}
}

\newglossaryentry{eq:v_pcc}
{
		name={\text{Voltage at reference bus}},
	description={xx}
}

\newglossaryentry{eq:u_min_bound}
{
		name={\text{lower bound on voltage magnitude}},
	description={xx}
}

\newglossaryentry{eq:u_max_bound}
{
		name={\text{Upper bound on voltage magnitude}},
	description={xx}
}

\newglossaryentry{eq:i_line_phase}
{
		name={\text{Line current definition}},
	description={xx}
}

\newglossaryentry{eq:i_line_max_bound}
{
		name={\text{Upper bound on line current}},
	description={xx}
}

\newglossaryentry{eq:s_injs}
{
		name={\text{Power injections}},
	description={xx}
}

\newglossaryentry{eq:s_set_constraints}
{
		name={\text{Complex power constraints}},
	description={xx}
}

\newglossaryentry{reb_cost}
{
		name={\text{Rebalancing cost}},
	description={xx}
}

\newglossaryentry{charger_cost}
{
		name={\text{Electricity cost, charging}},
	description={xx}
}

\newglossaryentry{losses_cost}
{
		name={\text{Electricity cost, losses}},
	description={xx}
}

\newglossaryentry{eamod_electricity_cost}
{
		name={\text{Electricity cost, \gls{amod}}},
	description={xx}
}

\newglossaryentry{eamod_total_cost}
{
		name={\text{Total cost, \gls{amod}}},
	description={xx}
}

\newglossaryentry{charger_energy}
{
		name={\text{Energy, charging}},
	description={xx}
}

\newglossaryentry{losses_energy}
{
		name={\text{Energy, losses}},
	description={xx}
}

\newglossaryentry{eamod_energy}
{
		name={\text{Energy, \gls{amod}}},
	description={xx}
}

\newglossaryentry{gan_lp}
{
		name={\text{\gls{bfm_lp}}},
	description={xx}
}

\newglossaryentry{gan_sdp}
{
		name={\text{\gls{bfm_sdp}}},
	description={xx}
}

\newglossaryentry{bolognani_lp}
{
		name={\text{\gls{lpfm_lp}}},
	description={xx}
}

\newglossaryentry{ml}
{
		name={\text{MATLAB}},
	description={xx}
}

\newglossaryentry{uot}
{
		name={Unbalanced OPF Toolkit},
	description={xx}
}

\newglossaryentry{yalmip}
{
		name={\text{YALMIP}},
	description={xx}
}

\newglossaryentry{oc}
{
		name={\text{Orange County, CA}},
	description={xx}
}

\newglossaryentry{pq_bus}
{
	name={\text{PQ-bus}},
	plural={\text{PQ-buses}},
	description={xx}
}

\newglossaryentry{matsim}
{
	name={\text{MATSim}},
	description={MATSim}
}

\begin{document}
\title{On the Interaction between \Glsentrylong{amod} Systems and Power Distribution Networks ---  An Optimal Power Flow Approach}

\author{Alvaro~Estandia*,
	    Maximilian~Schiffer,
	    Federico~Rossi,
	    Justin~Luke,
	    Emre Kara,
	    Ram Rajagopal,
      Marco~Pavone
\thanks{A. Estandia is with Marain Inc., Palo Alto, CA 94306, USA. He worked on this paper while he was a visiting student at Stanford University. Email: \texttt{alvaro@marain.com}.}
\thanks{M. Schiffer is with the TUM School of Management, Technical University of Munich, Munich 80333, Germany. Email: \texttt{schiffer@tum.de}.}%
\thanks{F. Rossi is with the NASA Jet Propulsion Laboratory, California Institute of Technology, Pasadena, CA 91109, USA. He worked on this paper while he was a Ph.D. student at Stanford University. Email: \texttt{federico.rossi@jpl.nasa.gov}.}%
\thanks{J. Luke is with the Department of Civil and Environmental Engineering, Stanford University, Stanford, CA 94035, USA. Email: \texttt{jthluke@stanford.edu}.}%
\thanks{E. Kara is with eIQ Mobility, Oakland, CA 94612, USA. He worked on this paper while he was with the SLAC National Accelerator Laboratory, Menlo Park, CA 94025, USA. Email: \texttt{eck@fastmail.com}.}%
\thanks{R. Rajagopal is with the Department of Civil and Environmental Engineering, Stanford University, Stanford, CA 94035, USA. Email: \texttt{ramr@stanford.edu}.}%
\thanks{M. Pavone is with the Department of Aeronautics and Astronautics, Stanford University, Stanford, CA 94035, USA.
	Email: \texttt{pavone@stanford.edu}.}
\thanks{*Corresponding author.}
\thanks{\textcopyright 2021 IEEE. Personal use of this material is permitted.  Permission from IEEE must be obtained for all other uses, in any current or future media, including reprinting/republishing this material for advertising or promotional purposes, creating new collective works, for resale or redistribution to servers or lists, or reuse of any copyrighted component of this work in other works.}
}

\maketitle

\begin{abstract}
In future transportation systems, the charging behavior of electric \gls{amod} fleets, i.e., fleets of electric self-driving cars that service on-demand trip requests, will likely challenge \glspl{pdn}, causing overloads or voltage drops.
In this paper, we show that these challenges can be significantly attenuated if the \glspl{pdn}' operational constraints and exogenous loads (e.g., from homes or businesses) are accounted for when operating an \gls{eamod} fleet.
We focus on a system-level perspective, assuming full coordination between the \gls{amod} and the \gls{pdn} operators. From this single entity perspective, we assess potential coordination benefits. Specifically, we extend previous results on an optimization-based modeling approach for \gls{eamod} systems to jointly control an \gls{eamod} fleet and a series of \glspl{pdn}, and analyze the benefit of coordination under load balancing constraints. For a case study of \gls{oc}, we show that the coordination between the \gls{eamod} fleet and the \glspl{pdn} eliminates 99\% of the overloads and 50\% of the voltage drops that the \gls{eamod} fleet would cause in an uncoordinated setting. Our results show that coordinating \gls{eamod} and \glspl{pdn} can help maintain the reliability of \glspl{pdn} under added \gls{eamod} charging load, thus significantly mitigating or deferring the need for \gls{pdn} capacity upgrades.
\glsresetall
\end{abstract}


\IEEEpeerreviewmaketitle

\section{Introduction}
\begin{figure}
	\centering
	\includegraphics[width=1.\linewidth]{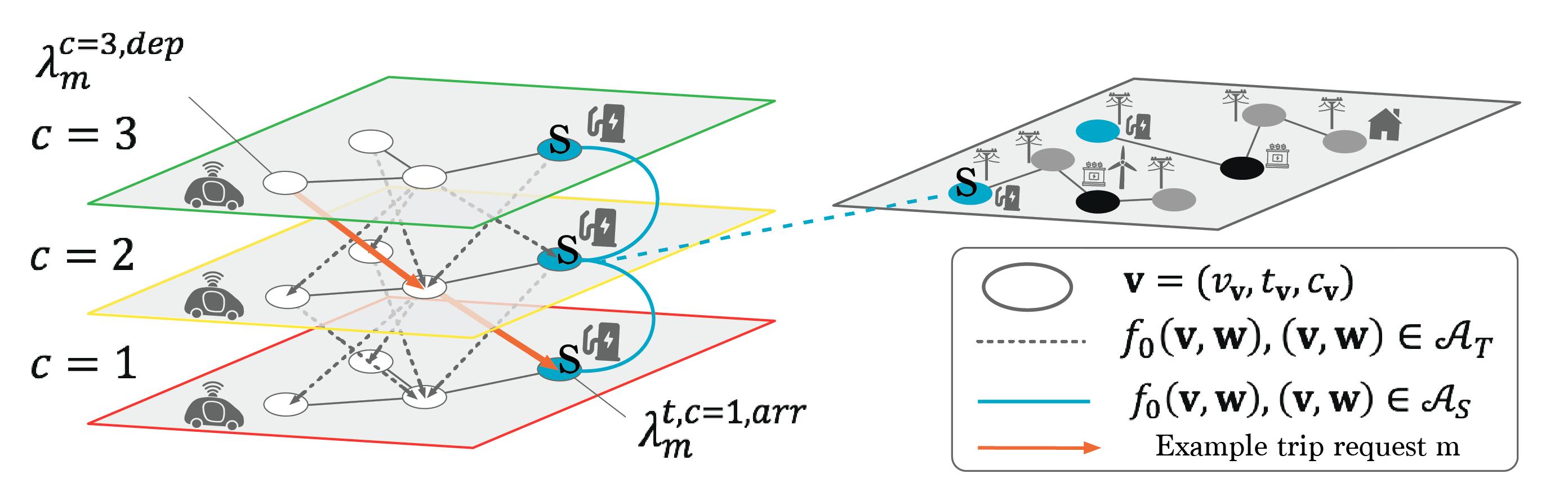}
	\caption{\label{fig:amod_power_interaction} Integration of an expanded road graph (left) and multiple \glspl{pdn} (right). Typically, a road network spans across multiple \glspl{pdn} and connects to the  \glspl{pdn} via charging station vertices. Besides charging stations that represent controllable loads, \glspl{pdn} contain reference buses (typically substations) highlighted in black and uncontrollable loads from residential and commercial customers.}
\end{figure}

\glsresetall
\IEEEPARstart{F}{leets} of electric self-driving cars servicing on-demand trip requests promise affordable urban mobility with reduced greenhouse gas emissions \citep{BauerGreenblattEtAl2018}, decreased need for parking \citep{SpieserTreleavenEtAl2014}, and fewer road accidents \citep{HannonMcKerracherEtAl2016}.
Additionally, such systems offer further benefits stemming from optimized central coordination, e.g., increased vehicle utilization compared to privately owned vehicles \citep{SpieserTreleavenEtAl2014}, and increased operational flexibility and efficiency compared to taxi, car-sharing, and ride-hailing services.
Furthermore, \gls{eamod} has the potential to foster the adoption of \glspl{ev}
since, in a high-utilization fleet, \glspl{ev} are more economical than their gasoline-powered counterparts~\citep{BauerGreenblattEtAl2018}.
Nonetheless, operating an \gls{eamod} fleet also bears inherent challenges as \glspl{ev} show range limitations which require time-consuming recharging that adds a sizable load on \glspl{pdn}.
Studies on private \glspl{ev} showed that uncoordinated charging may require costly \gls{pdn} upgrades to secure stabilization, as it can destabilize \glspl{pdn} due to overloads or under-voltages \citep{ShareefIslamEtAl2016,RichardsonFlynnEtAl2012,Clement-NynsHaesenEtAl2010}.
In contrast, we expect that intelligently coordinating the vehicles' charging would reduce such negative impacts, in particular, by reducing or deferring the need for power network upgrades.

Controlling \gls{amod} systems entails solving a dispatching problem to assign vehicles to on-demand trip requests.
The system's performance increases if empty vehicles are proactively repositioned (rebalanced) in anticipation of future demand \citep{PavoneSmithEtAl2012}.
In the past decade, multiple approaches have been presented for the control of \gls{amod} systems with varying degrees of mathematical complexity.
A first family of algorithms relies on heuristic rules to dispatch and rebalance a fleet \citep{BischoffMaciejewski2016,FagnantKockelman2014}.
More sophisticated methods use optimization algorithms to control the \gls{amod} system.
Often, network flow models using fluidic relaxations, i.e., allowing for fractional vehicles and fractionally serviced trip requests are used \citep{PavoneSmithEtAl2012}.
Models of this type have been extended to consider road capacities and congestion \citep{RossiZhangEtAl2017}.

To control an \emph{electric} \gls{amod} system, an operator must keep track of a vehicle's \gls{soc} and recharge a vehicle's battery accordingly.
Again, some heuristic approaches exist \citep{ChenKockelmanEtAl2016,Munkhammar2017}.
Optimization-based algorithms are so far not amenable to large-scale problems as they rely on \glspl{milp} with discretized \glspl{soc} \citep{ZhangRossiEtAl2016b,Dandl2018}. 

At its core, the operation of an \gls{eamod} system induces a coupling between the power network and the transportation system.
Specifically, the \gls{eamod} fleet represents a controllable load in time and space.
All previously mentioned studies neglect the impact of an \gls{eamod} system on the power grid, despite the fact that even a moderate amount of \glspl{ev} may significantly increase electricity prices \citep{HadleyTsvetkova2009} and may negatively influence the power grid's reliability \citep{TranBanisterEtAl2012}, \citep{Veldman2015}.
A few recent studies consider such a coupling {\em implicitly} via available capacities \citep{Yu2018} or prices \citep{Iacobucci2018}, but the proposed control algorithms for the \gls{eamod} fleet do not explicitly account for the fleet impact on the power network. Only \citet{RossiIglesiasEtAl2018b} consider the fleet impact on the power network explicitly, introducing the \gls{pamod} model, a linear model that combines a network flow model for the \gls{eamod} system and a balanced single-phase DC model of a \emph{transmission} network.
However, this model does not consider the power distribution network, which is the more appropriate grid stage to analyze mesoscopic \glspl{ev} fleet operations \citep{Leou2014}. Notably, a single-phase DC model is not sufficient to model a \gls{pdn} as it assumes a constant voltage magnitude, and neglects reactive power and link resistances \citep{GanLow2014}; instead, a three-phase model is necessary \citep[Ch. 1]{Kersting2002}. So far, \glspl{pdn} were only considered when determining optimal charging schedules for privately owned \glspl{ev} which have to reach a certain \gls{soc} by the end of a given planning horizon \citep{Clement-NynsHaesenEtAl2010,RichardsonFlynnEtAl2012,HoogAlpcanEtAl2015} as opposed to centrally coordinated fleets.
Here, an instance of the \acrfull{opf} problem can be solved to balance necessary charging loads with \gls{pdn}-specific constraints.

In summary, individual aspects of the control problem addressed in this paper, such as the control of an \gls{eamod} system or considering \gls{pdn} models to optimally charge private \glspl{ev}, have been addressed in the literature.
However, to the best of our knowledge, no studies that tightly couple an  \gls{eamod} system and \gls{pdn} models currently exist.

This work addresses this gap.
Specifically, our contribution is threefold. First, we present a benchmark of convex three-phase \gls{pdn} power flow approximations and identify a model compatible with the characteristics of the \gls{eamod} problem. We then extend the mesoscopic model in \cite{RossiIglesiasEtAl2018b} to capture the operations of and interaction between an \gls{eamod} system and a series of unbalanced \glspl{pdn}.
Second, we embed this model within an optimization problem that assesses achievable benefits with respect to full cooperation between the two systems. The mesoscopic optimization's solution enables comprehensive analyses to identify bottlenecks in \glspl{pdn} and inform operator decisions in the day-ahead electricity market.
Third, we provide a case study of \gls{oc} where we study the impact of an \gls{eamod} system on
the \glspl{pdn} and evaluate the benefits of coordination.

The remainder of this paper is structured as follows: \cref{sec:eamod} reports the mesoscopic model for an \gls{eamod} system used in previous work for self-consistency.  \cref{sec:powermodel} surveys existing \gls{pdn} models and identifies a suitable model for the \gls{eamod} application.
\cref{sub:amod_opf} discusses the interaction between the \gls{eamod} system and a series of \glspl{pdn}.
\Cref{sec:amod_opf_case_study} details our case study of \gls{oc}, and presents results that characterize the impact of \gls{eamod} systems on \glspl{pdn}, highlighting the improvement potential stemming from coordination.
\Cref{sec:conclusion} concludes this paper with a summary of its main findings and an outlook on future research.
Finally, the Appendix summarizes our notation and nomenclature.

\section{Modeling electric AMoD systems}
\label{sec:eamod}
In an \gls{amod} system, a fleet of autonomous vehicles services customer transportation requests, i.e., picks up customers at their origin and brings them to their destination \citep{SpieserTreleavenEtAl2014}.
A fleet operator controls the \gls{amod} fleet by assigning vehicles to customer requests and by routing each vehicle. 
Besides origin-destination trips of customers, the routing may comprise rebalancing trips in-between two customer trips as spatial and temporal mismatches between origins and destinations of different customer requests arise. 
In an \gls{eamod} problem, the fleet operator additionally controls vehicle charging schedules and rebalances vehicles based on anticipated spatial-temporal variations of vehicle \glspl{soc} and electricity prices.

We model an \gls{eamod} system with a network flow model as originally presented in \citep{RossiIglesiasEtAl2018b}, reported in this section for self-consistency. Sections~\ref{sec:powermodel} and \ref{sub:amod_opf} then detail our main contribution by integrating this model with \glspl{pdn}.
To avoid integer variables, the model uses \textit{i}) a fluidic vehicle approximation and \textit{ii})
a road graph expanded along two dimensions: discrete-time and vehicles' \gls{soc}. 

\paragraph{General road network representation}\mbox{} We model the road network as a graph $ G_{R} = (\gls{road_node_set},\gls{road_edge_set})$ with a set of vertices $\gls{in_road_node} \in \gls{road_node_set}$ and a set of road segment arcs $(\gls{in_road_edge}) \in \gls{road_edge_set}$. Each arc $(\gls{in_road_edge}) \in \gls{road_edge_set}$ is characterized by a distance $\gls{d_edge}$, a traversal time $\gls{t_edge}$, and an energy consumption $\gls{c_edge}$.

We consider a set $\gls{t_set} = \arange{\gls{n_t}}$ of discrete equidistant time steps (each of duration $\gls{t_step} \in \gls{reals_plus}$), and a set $\gls{c_set} = \arange{\gls{n_c}}$ of equidistant discrete battery charge levels (each of energy $\gls{c_unit} \in \gls{reals_plus}$).

While some vertices in $G_{R}$ merely represent intersections or access points, others represent charging stations $\gls{charger_set} \subseteq \gls{road_node_set}$ that allow recharging of vehicles. 
Each charging station $\gls{in_charger}\in\gls{charger_set}$ has a charging rate $\gls{charge_rate_charger} \in \arange{\gls{n_c}}$ that denotes the amount of \gls{soc} that can be recharged in a single time step.
Additionally, charging stations have a certain number of charging plugs $\gls{charger_cap} \in \mathbb{N}^+$ which limits the number of concurrently charging vehicles. 

We model congestion using a threshold model, i.e., we assume that vehicles drive at the road's free-flow speed as long as their number is less than the road's capacity $\gls{f_edge_cap} \in \gls{reals_plus}$, as detailed in \cite{RossiIglesiasEtAl2018b}.

\paragraph{Expanded graph representation}\mbox{} 
We use an expanded graph to model a vehicle's location and \gls{soc} over time. 
The expanded graph $\gls{exp_network} = (\gls{exp_node_set},\gls{exp_edge_set})$ is directed and has a vertex set $\gls{exp_node_set} \subseteq \gls{road_node_set}\times\gls{t_set}\times \gls{c_set}$. 
Each vertex $\gls{in_exp_node}\in \gls{exp_node_set}$ is defined by a tuple $(\gls{in_exp_node_road},\, \gls{in_exp_node_t},\, \gls{in_exp_node_c})$ that represents a vertex \gls{in_exp_node_road} of $\gls{road_node_set}$ at a specific time \gls{in_exp_node_t} with a specific \gls{soc} \gls{in_exp_node_c}. Figure~\ref{fig:amod_power_interaction} (left) illustrates the concept of \gls{soc} expansion; for ease of representation, the time expansion is not shown.
The resulting arc set \gls{exp_edge_set} consists of two subsets $\gls{exp_edge_set_road} \cup \gls{exp_edge_set_charger} = \gls{exp_edge_set}$.
Arcs $(\gls{in_exp_edge})\in \gls{exp_edge_set_road}$ represent travel in the road network and must meet the following condition
\begin{equation*}
\begin{split}
\gls{exp_edge_set_road} = &\lbrace (\gls{in_exp_edge}) \in \gls{exp_edge_set}\mid (\gls{in_exp_edge_road}) \in \gls{road_edge_set},\\
& \gls{in_exp_node_t_post} - \gls{in_exp_node_t} 	= \gls{t}_{\gls{in_exp_edge_road}},\, \gls{in_exp_node_c} -\gls{in_exp_node_c_post} = \gls{c}_{\gls{in_exp_edge_road}} \rbrace,
\end{split}
\end{equation*}
i.e.: \textit{i}) $(\gls{in_exp_edge_road})$ is a road arc, \textit{ii}) the time expansion $\gls{in_exp_node_t_post} - \gls{in_exp_node_t}$ equals its traversal time $\gls{t}_{\gls{in_exp_edge_road}}$, and \textit{iii}) the \gls{soc} expansion $\gls{in_exp_node_c_post} - \gls{in_exp_node_c}$ equals its consumption $\gls{c}_{\gls{in_exp_edge_road}}$.
Arcs $(\gls{in_exp_edge})\in \gls{exp_edge_set_charger}$ represent 
recharging
at a charging station and must meet the following condition
\begin{equation*}
\begin{split}
\gls{exp_edge_set_charger}\! = & \lbrace (\gls{in_exp_edge}) \!\in\! \gls{exp_edge_set} \mid \gls{in_exp_node_road}\!=\!\gls{in_exp_node_road_post} \!=\! \gls{in_charger} \!\in\! \gls{charger_set},
\gls{in_exp_node_c_post}\! -\! \gls{in_exp_node_c}\! =\! (\gls{in_exp_node_t_post} - \gls{in_exp_node_t})  \gls{charge_rate_charger}
\rbrace,
\end{split}
\end{equation*}
i.e.: \textit{i}) \gls{in_exp_node_road} and \gls{in_exp_node_road_post} are equal and correspond to a charging station, and \textit{ii}) the \gls{soc} difference $\gls{in_exp_node_c_post} - \gls{in_exp_node_c}$ equals the amount of energy recharged, that is $(\gls{in_exp_node_t_post} - \gls{in_exp_node_t})  \gls{charge_rate_charger}$.
\paragraph{Customer trip requests}\mbox{} 
In addition to this graph representation, we define a \glsdesc{m_set} $\gls{m_set} = \arange{\gls{n_m}}$. 
Each trip $\gls{in_m} \in \gls{m_set}$ is defined by a quadruple $(\gls{origin_node_m},\,\gls{dest_node_m},\,\gls{t_m},\,\gls{trip_rate_m}) \in \gls{road_node_set} \times \gls{road_node_set} \times \gls{t_set} \times \gls{reals_plus}$ that denotes its origin $\gls{in_road_node}_{\gls{in_m}}$, its destination $\gls{in_road_node_post}_{\gls{in_m}}$, its departure timestep $\gls{t}_{\gls{in_m}}$, and the number of customer trip requests (i.e., the number of customers who wish to travel between $\gls{in_road_node}_{\gls{in_m}}$ and $\gls{in_road_node_post}_{\gls{in_m}}$ departing at $\gls{t}_{\gls{in_m}}$) $\gls{trip_rate_m}$.
We assume a deterministic setting in which these requests are known or predicted for all timesteps.
To reduce the number of decision variables, we use precomputed vehicle routes for customer-carrying vehicles, corresponding to shortest-time paths $\gls{route_precomp}$ that do not violate the congestion constraints. As we use a threshold congestion model, we can straightforwardly precompute such feasible shortest-time paths by solving a network flow problem as in \cite{RossiIglesiasEtAl2018b}. Each shortest-time path has a traveling time $\gls{route_precomp_t}$ and a charge requirement $\gls{route_precomp_c}$.
We denote \gls{trip_rate_in_c} as the \glsdesc{trip_rate_in_c} and \gls{trip_rate_out_t_c} as the \glsdesc{trip_rate_out_t_c}.
Thus,
\begin{equation}
\label{eq:customer_charge_conservation}
\begin{split}
\gls{trip_rate}_{\gls{in_m}}^{\gls{t},\gls{c},\gls{out}} = 
\begin{cases}
\gls{trip_rate}_{\gls{in_m}}^{\gls{c} + \gls{c}_{\gls{in_road_node}_{\gls{in_m}}\to\gls{in_road_node_post}_{\gls{in_m}}},\gls{in}} & \text{ if } \gls{t}_{\gls{in_m}} = \gls{t} - \gls{t}_{\gls{in_road_node}_{\gls{in_m}}\to\gls{in_road_node_post}_{\gls{in_m}}} \\
0 & \text{otherwise}
\end{cases}\\
\hfill \forall \gls{t} \in \gls{t_set},\, \gls{c} \in \gls{c_set},\,
\forall\gls{in_m} \in \gls{m_set}.
\end{split}
\end{equation}
\paragraph{Electric AMoD model}\mbox{} 
We introduce $\gls{f_0}(\gls{in_exp_edge}) : \gls{exp_edge_set} \to \gls{reals_plus}$  to represent the flow of customer-empty vehicles on arc $(\gls{in_exp_edge})$, which includes both rebalancing and charging vehicles.  
Further, $\gls{dist_vehicle_initial}(\gls{in_exp_node})$ denotes the initial location of the vehicles, i.e., the number of vehicles available at vertex $\gls{in_exp_node_road}$ with charge level \gls{in_exp_node_c} at $\gls{in_exp_node_t} = 1$ and is zero for all other timesteps. 
Analogously, $\gls{dist_vehicle_final}(\gls{in_exp_node})$
denotes the desired final location of the vehicles, i.e., the number of vehicles that must be at node $\gls{in_exp_node_road}$ with charge level \gls{in_exp_node_c} at $\gls{in_exp_node_t} = \gls{n_t}$.
With this notation, a multi-commodity flow representation of the \gls{eamod} model is given by:
\begin{align}
\label{eq:reb_conservation}
&\csum{\gls{in_exp_node_post}:(\gls{in_exp_edge})\in\gls{exp_edge_set}} \gls{f_0}(\gls{in_exp_edge}) + \csum{\gls{in_m}=1}^{\gls{n_m}} \gls{indicator}_{\gls{in_exp_node_road} = \gls{in_road_node}_{\gls{in_m}}} \gls{indicator}_{\gls{in_exp_node_t} = \gls{t}_{\gls{in_m}}} \gls{trip_rate}_{\gls{in_m}}^{\gls{in_exp_node_c},\gls{in}} + \gls{dist_vehicle_final}(\gls{in_exp_node}) \\
&= \csum{\gls{in_exp_node_pre}:(\gls{in_exp_edge_pre})\in\gls{exp_edge_set}} \gls{f_0}(\gls{in_exp_edge_pre}) + \csum{\gls{in_m}=1}^{\gls{n_m}} \gls{indicator}_{\gls{in_exp_node_road} = \gls{in_road_node_post}_{\gls{in_m}}} \gls{trip_rate}_{\gls{in_m}}^{\gls{in_exp_node_t},\gls{in_exp_node_c},\gls{out}} + \gls{dist_vehicle_initial}(\gls{in_exp_node}) \quad \forall\gls{in_exp_node} \in \gls{exp_node_set}, \nonumber
\end{align}
\begin{align}
\sum_{\gls{c} = 1}^{\gls{n_c}}\gls{trip_rate_in_c} \!=\! \gls{trip_rate_m}  \, \forall \gls{in_m} \!\in\! \gls{m_set},
\,\,
\sum_{\gls{t} = 1}^{\gls{n_t}} \sum_{\gls{c} = 1}^{\gls{n_c}} \gls{trip_rate_out_t_c} \!=\! \gls{trip_rate_m} \, \forall \gls{in_m} \!\in\! \gls{m_set}  \label{eq:source_sink_conservation}
\end{align}
Here, $\gls{indicator}_x$ is the indicator function.
\Cref{eq:reb_conservation} secures flow conservation for rebalancing and charging vehicles,
ensures a sufficient number of empty vehicles in each vertex to cover originating trip requests, and enforces initial and final conditions on the vehicle locations through \gls{dist_vehicle_initial} and \gls{dist_vehicle_final}.
\Cref{eq:source_sink_conservation} distributes the demand for a given trip request $\gls{in_m}$ to vehicles with different \gls{soc}, and accumulates vehicles arriving at different times with different \gls{soc} for request $\gls{in_m}$. 

\paragraph{Electric AMoD problem}\mbox{}
\label{par:eamod_problem}
We now extend the basic constraints of the \gls{eamod} model to a full \gls{eamod} model.
Specifically, we optimize the vehicles' rebalancing routes and charging schedules in order to minimize the cost of operating the \gls{eamod} system, that is:
\begin{mini!}
	{\substack{\gls{f_0},[\gls{trip_rate_in_c}]_{\gls{c} \in \gls{c_set}}, \\ [\gls{trip_rate_out_t_c}]_{\gls{c} \in \gls{c_set},\gls{t}\in\gls{t_set}}, \gls{dist_vehicle_initial}, \gls{dist_vehicle_final}}}
	{\gls{distance_price}\csum{(\gls{in_exp_edge}) \in \gls{exp_edge_set_road}} \gls{d}_{\gls{in_exp_edge_road}} \gls{f_0}(\gls{in_exp_edge})\nonumber}
	{\label{eq:eamod_problem}}{}
	\breakObjective {+ 
		\csum{\substack{(\gls{in_exp_edge}) \in \gls{exp_edge_set_charger}:\gls{in_exp_node_road} = \gls{in_exp_node_road_post} = \gls{in_charger}}} \gls{price}_{\textrm{el},\gls{in_charger}}[\gls{in_exp_node_t}] \gls{charge_rate_charger} \gls{f_0}(\gls{in_exp_edge})\label{eq:eamod_problem_obj}}
	\addConstraint{
	\text{\cref{eq:reb_conservation,eq:source_sink_conservation,eq:customer_charge_conservation}}
	}{\quad}{\text{\Gls{eamod} model}\nonumber}
		\addConstraint{
		\csum{\substack{(\gls{in_exp_edge}) \in \gls{exp_edge_set_road}:\\\gls{in_exp_node_road} = \gls{in_road_node}, \gls{in_exp_node_road_post} = \gls{in_road_node_post}, \gls{in_exp_node_t} = \gls{t}}} \gls{f_0}(\gls{in_exp_edge})}{\leq \gls{f_exp_edge_cap_t}\quad}{\forall (\gls{in_road_edge}) \in \gls{road_edge_set},\, \gls{t} \in \gls{t_set}
		\label{eq:road_capacity_adjusted}}	
	\addConstraint{
		\csum{\substack{(\gls{in_exp_edge}) \in \gls{exp_edge_set_charger}:\\\gls{in_exp_node_road} = \gls{in_exp_node_road_post} = \gls{in_charger}, \gls{in_exp_node_t} = \gls{t}}}
		\gls{f_0}(\gls{in_exp_edge})}{\leq \gls{charger_cap}}{\forall\gls{in_charger} \in \gls{charger_set},\, \gls{t} \in \gls{t_set}
		\label{eq:charger_capacity_physical}}	
	\addConstraint{g_{I}(\gls{dist_vehicle_initial})}{= 0}{g_{F}(\gls{dist_vehicle_final})= 0}  {\label{eq:dist_vehicle_initial}}
\end{mini!}
Here, we use the previously introduced concept of expanded graph vertices: each vertex $\gls{in_exp_node}\in \gls{exp_node_set}$ is defined by a tuple $(\gls{in_exp_node_road},\, \gls{in_exp_node_t},\, \gls{in_exp_node_c})\in \gls{road_node_set}\times\gls{t_set}\times \gls{c_set}$. The objective function \cref{eq:eamod_problem_obj}  minimizes the operational cost of the \gls{eamod} system, considering time-invariant operational cost per unit distance (e.g., discounted cost for maintenance, tires, depreciation) $\gls{distance_price} \in \gls{reals}$ for rebalancing vehicles and time-varying electricity costs $\gls{electricity_price_charger} \in \gls{reals}$ for recharging vehicles at a charging station $\gls{in_charger} \in \gls{charger_set}$. Figure~\ref{fig:amod_power_interaction} depicts example arcs that model such rebalancing and charging flows (\gls{f_0}), as well as \gls{trip_rate_in_c} and \gls{trip_rate_out_t_c} for an example trip \gls{in_m} marked with bold arrows.
\Cref{eq:reb_conservation,eq:source_sink_conservation,eq:customer_charge_conservation}
impose general flow conservation while \cref{eq:road_capacity_adjusted} applies the threshold congestion model to 
rebalancing flows. 
As customer-carrying flows are fixed, we do not consider these directly in \cref{eq:road_capacity_adjusted}.
Instead, we use the \glsdesc{f_exp_edge_cap_t} $\gls{f_exp_edge_cap_t}$ which results from subtracting the customer carrying flow on road arc $(\gls{in_road_edge})$ at time step $\gls{t}$ from the corresponding \glsdesc{f_cap} $\gls{f_edge_cap}$.
The pre-routed vehicles may congest a road link. 
In this case, we set the residual capacity $\gls{f_exp_edge_cap_t}$ for that link to zero. Thus, customer-carrying flows and residual capacity are fixed and constant with respect to the optimization of rebalancing flows.
\Cref{eq:charger_capacity_physical} limits the number of vehicles that can use a charging station concurrently according to the number of charging plugs at each station. 
We impose initial and final conditions on vehicles with the generic functions $g_{I}$ and $g_{F}$ in \cref{eq:dist_vehicle_initial}.
The brackets in the decision variables denote concatenation. 
We will use this convention in the rest of the paper.

The \gls{eamod} problem \cref{eq:eamod_problem} has $\gls{n_t}\gls{n_c}(|\gls{road_edge_set}| + |\gls{charger_set}|) + \gls{n_c}\gls{n_m} + \gls{n_t}\gls{n_c}|\gls{road_node_set}| + \gls{n_c}|\gls{road_node_set}|$ decision variables. 
The dominant term is $\gls{n_c}\gls{n_m}$: there could be at most one customer trip request from every origin to every destination at every time step such that $\gls{n_m} \leq |\gls{road_node_set}|^2 \gls{n_t}$. 
It follows that $\gls{n_c}\gls{n_m} \in \gls{bigO}(\gls{n_c}|\gls{road_node_set}|^2\gls{n_t}).$

A few comments are in order.
First, we consider discrete time steps as well as discrete \gls{soc} levels. From a mesoscopic viewpoint, these discretizations bear sufficient accuracy while improving the model's computational tractability significantly.
Second, the network flow model treats vehicles and customers as fractional flows; accordingly, it is not readily suitable for real-time control of \gls{eamod} fleets. 
Again, this accuracy loss is acceptable at a mesoscopic level and is compatible with our goal of assessing the achievable performance stemming from the coordination between \gls{eamod} and \gls{pdn} operators. 
Note that our solution can still be used as a reference plan for a lower-level microscopic controller \citep[cf.][]{IacobucciMcLellanEtAl2019}.
Third, we limit the vehicle flow on a given road link to its capacity and assume vehicles travel at free-flow speed accordingly. Such a threshold congestion model is in line with the accuracy requirements of our mesoscopic viewpoint. If necessary,
more sophisticated congestion models can easily be integrated into our modeling approach, at the cost of computational tractability.
Fourth, our model does not explicitly account for congestion from non-\gls{amod} traffic. 
However, this type of traffic can be considered by subtracting the corresponding flow from the \glsdesc{f_exp_edge_cap_t} $\gls{f_exp_edge_cap_t}$.
Fifth, we assume that future trip requests are known or estimated with a high degree of accuracy. 
While the development of tools to estimate \gls{amod} demand is beyond the scope of this paper, remarkably accurate algorithms are available in the literature \citep[e.g.,][]{IglesiasRossiEtAl2018}.
Sixth, we optimize only rebalancing trips and fix customer trips to their shortest-time-paths.
In principle, including the optimization of customer-carrying trips could yield solutions with lower cost;
however, our prior work has shown that the inclusion of customer-carrying trips in the optimization problem results in a small decrease in cost at the price of a significant increase in computational complexity \citep{RossiIglesiasEtAl2018b}.
Also, note that although the route of customer-carrying trips is fixed, the \gls{soc} of customer-carrying vehicles is part of the optimization problem.
Finally, the \gls{eamod} problem \cref{eq:eamod_problem} may become infeasible if the number or the distribution of customer trip requests exceeds the customer-carrying capacity of the \gls{eamod} system.
Here, we assume that the problem is always feasible as the fleet operator can reject or postpone trip requests to ensure feasibility. This is in line with common practice in today's taxi or ride-hailing platforms.
Nonetheless, a mechanism to decide which trips should be rejected or postponed is beyond the scope of this paper.

\section{Modeling unbalanced \glsentrylongpl{pdn}}
\label{sec:powermodel}
This section provides the basics for modeling unbalanced \glspl{pdn} and presents the identification of a compatible convex power flow surrogate to model the integration of \gls{pdn} into an \gls{eamod} model under a unified notation framework. 
First, we introduce an unbalanced \gls{pdn} model in \cref{sub:powermodel}.
Then, we define the optimal power flow problem in \cref{sec:opf_formulation}.
Finally, we compare convex power flow surrogates in \cref{sub:power_flow_surrogates} and justify the selected surrogate.

\jvspace{-1em}
\subsection{Unbalanced \glsentrylong{pdn} model}
\label{sub:powermodel}
In the following, we consider only radial network structures which is the typical configuration for \glspl{pdn} \cite[Ch. 1.1]{Kersting2002} and base our notation on \citep{GanLow2014}.
A radial \gls{pdn} is modeled as a directed graph $\gls{dist_net} = (\gls{bus_set},\gls{line_set})$ with a tree topology, consisting of a \glsdesc{bus_set} $\gls{bus_set} = \arange[0]{\gls{n_bus}}$ 
and a \glsdesc{line_set} $\gls{line_set} \subset \gls{bus_set}^2$.

Each \gls{pdn} has a reference bus which typically denotes a substation that connects the \gls{pdn} to the transmission network.
The set $\gls{bus_set_plus} \!=\! \gls{bus_set}\setminus 0$ contains all buses other than the reference bus 0.
Buses are connected by links (e.g., power lines, transformers, regulators), such that $(\gls{in_line}) \in \gls{line_set}$ represents a link between $\gls{in_bus}$ and $\gls{in_bus_post}$
for which \gls{in_bus} lies in the single path between the reference bus 0 and bus \gls{in_bus_post}.
Note that there is only one such path because, by assumption, $\gls{dist_net}$ is a tree.

We consider unbalanced \glspl{pdn} with three phases $\gls{phase_set} = \lbrace a, b, c \rbrace$.
In line with this, $\gls{phase_set_line} \subseteq \gls{phase_set}$ is the \glsdesc{phase_set_line}.
Further, the \glsdesc{phase_set_bus} comprises the phases of all links connected to the bus:
\begin{equation*}
\gls{phase_set_bus} =  \left(\cup_{(\gls{in_line_pre}) \in \gls{line_set}} \gls{phase_set}_{\gls{in_line_pre}}\right) \cup \left(\cup_{(\gls{in_line}) \in \gls{line_set}} \gls{phase_set_line} \right)  \quad \forall \gls{in_bus} \in \gls{bus_set}.
\end{equation*}

Each bus \gls{in_bus} has a time-invariant \glsdesc{y_mat_shunt} $\gls{y_mat_shunt_bus} \in \gls{complexes}^{\gls{n_phase_set_bus} \times \gls{n_phase_set_bus}}$, representing the admittance between the bus and ground.
Further, each link $(\gls{in_line})$ has a time-invariant \glsdesc{z_mat} $\gls{z_mat_line} \in \gls{complexes}^{\myabs{\gls{phase_set}_{\gls{in_line}}} \times \myabs{\gls{phase_set}_{\gls{in_line}}}}$.

We consider a discrete-time model that tracks a series of steady states in the power network and neglects dynamic effects. This is appropriate if the discretization time is substantially longer than the time scale for the dynamic effects (i.e., in the order of minutes). 
We consider a timespan $\gls{t_set} = \arange{\gls{n_t}}$ with time steps $t\in\gls{t_set}$, each having a length $\gls{t_step} \in \gls{reals_plus}$. 
Each bus \gls{in_bus} has a time-dependent \glsdesc{v} $\gls{v_bus_phase}\foft  \in \gls{complexes}$ and a \glsdesc{s_inj}  $\gls{s_inj_bus_phase}\foft \in \gls{complexes}$ for each of its phases. 
Concurrently, each link shows a time-dependent current for each of its phases $\gls{i_line_phase}\foft \in \gls{complexes}$. 
For brevity, we use vectors for per-phase quantities: ${\gls{v_bus} = [\gls{v_bus_phase}]_{\gls{in_phase} \in \gls{phase_set_bus}}}$, 
${\gls{s_inj_bus} = [\gls{s_inj_bus_phase}]_{\gls{in_phase} \in \gls{phase_set_bus}}}$, and
${\gls{i_line} = [\gls{i_line_phase}]_{\gls{in_phase} \in \gls{phase_set_line}}}$.
Herein, superscripts represent the projection onto specific phases.

The current on each link obeys Ohm's law, that is:
\begin{equation*}
\gls{i_line_phase}\foft = \gls{y_mat}_{\gls{in_line}}((\gls{v_vec}_{\gls{in_bus}}\foft)^{\gls{phase_set_line}} - (\gls{v_vec}_{\gls{in_bus_post}}\foft)^{\gls{phase_set_line}}) \quad (\gls{in_line}) \in \gls{line_set},\, \gls{t} \in \gls{t_set},
\end{equation*} 
with $\gls{y_mat}_{\gls{in_line}} = \gls{z_mat}_{\gls{in_line}}^{-1}$ \citep{GanLow2014}.
Each bus is either specified by its voltage or by its power injection such that the remaining quantity is a dependent variable \citep[Ch. 6.4]{GloverSarmaEtAl2011}.
We refer to specified variables as \emph{direct variables} and to those that are dependent as \emph{indirect variables}.
The reference bus specifies the reference voltage $\gls{v_ref_phase}\foft \in \gls{reals}$ for the network:
\begin{equation}
\gls{v_pcc_phase}\foft = \gls{v_ref_phase}\foft \quad \gls{in_phase} \in \gls{phase_set}_{\gls{pcc}},\, \gls{t} \in \gls{t_set}.  
\label{eq:v_pcc}
\end{equation} 
Accordingly, the \glsdesc{v} $\gls{v_vec}_{0}$ is the direct variable and the \glsdesc{s_inj} $\vec{s}_{\gls{inj},0}$ remains dependent.

For all other buses $\gls{in_bus}\in \gls{bus_set_plus}$, the \glsdesc{s_inj} \gls{s_inj_bus} is the direct variable, whereas the \glsdesc{v} \gls{v_bus} remains dependent.
These buses are called PQ buses since the active ($p$) and reactive power injection ($q$) are the direct variables.
Herein, each \gls{pq_bus} has a time-varying uncontrollable load with complex power ${\gls{unc_load_s_bus}\foft \in \gls{complexes}^{\gls{n_phase_set_bus}}}$. 
These loads represent electricity demand from residential and commercial customers.
We consider uncontrollable loads to be exogenous but known in advance within timespan $\gls{t_set}$. 

Controllable loads $\gls{in_con_load} \in \gls{con_load_set} = \arange[1]{\gls{n_con_load}}$ are defined by a tuple ${(\gls{con_load_s_c_vec}\foft,\gls{con_load_bus}) \in \gls{complexes}^{|\gls{phase_set}_{\gls{con_load_bus}}|}\times\gls{bus_set}}$ denoting their complex power $\gls{con_load_s_c_vec}$ and its corresponding bus \gls{con_load_bus}. 
These loads represent dispatchable generators or loads that can be throttled.
With this notation, the power injections at \glspl{pq_bus} are
{ 
\begin{equation}
\label{eq:s_inj_n}
\gls{s_inj_bus}\foft \!\!=\! - \gls{unc_load_s_bus}\foft \!-\! \sum_{\gls{in_con_load} = 1}^{\gls{n_con_load}} \gls{indicator}_{n = \gls{con_load_bus}} \gls{con_load_s_c_vec}\foft \,\, \gls{in_bus} \in \gls{bus_set_plus},\gls{t} \in \gls{t_set}.
\end{equation}
}
Note that we model generators as negative loads without loss of generality. 
Further, we consider only wye-connected constant power loads which may require performing delta-to-wye conversions for some loads and approximating constant current and constant impedance loads as constant power ones.
This simplification is common in optimization frameworks~\citep{BernsteinDallAnese2017}.

Dependent variables result from the network topology and its controllable and uncontrollable loads.
Specifically, they are related by the power flow equation \citep{ZhaoDallAneseEtAl2017}
\begin{align}
\label{eq:powerflow}
&\gls{s_inj_bus}\foft 
= \diag{\gls{v_bus}\foft\gls{v_bus}\foft\H\gls{y_mat_shunt}_{\gls{in_bus}}\H} \\
&+ \csum{\gls{in_bus}:(\gls{in_line})\in\gls{line_set}} \diag{\gls{v_bus}^{\gls{phase_set_line}}\foft(\gls{v_bus}^{\gls{phase_set_line}}\foft - \gls{v}_{\gls{in_bus_post}}^{\gls{phase_set_line}}\foft)\H\gls{y_mat}_{\gls{in_line}}\H}^{\gls{phase_set_bus}} \quad \gls{t} \in \gls{t_set}. \nonumber 
\end{align}

Collectively, these equations allow us to model a radial time-invariant unbalanced \gls{pdn} with time-varying controllable and uncontrollable loads.

A few comments are in order.
First, we consider a discrete-time model that tracks a series of steady states in the power network. As we are not interested in dynamic effects, this model is appropriate, and the level of aggregation is aligned with our mesoscopic transportation model.
Second, we consider a time-invariant \gls{pdn} which cannot model control elements, e.g., step voltage regulators. 
Optimization frameworks commonly neglect these elements (see \citep[][]{HoogAlpcanEtAl2015,ZhaoDallAneseEtAl2017}) as their inclusion substantially increases complexity while their omission results in a more conservative optimization. This simplification is appropriate for the purposes of a mesoscopic system-level analysis.
Third, we assume that high-quality estimates of uncontrollable electrical loads are available. 
While deriving such estimates exceeds our scope, techniques to accurately estimate future power demand exist~\citep[e.g.,][]{HernandezBaladronEtAl2014}.

\jvspace{-1em}
\subsection{Optimal Power Flow problem}
\label{sec:opf_formulation}
The \gls{opf} problem \cref{eq:opf_problem} optimizes a  power network's state subject to its operational constraints and is often used to support grid-related decisions, e.g.,
operational or strategic planning, and pricing \citep{Taylor2015}.
Here, we use an \gls{opf} problem for operational planning and decide on the controllable loads while optimizing a generic objective function $\gls{obj_fun}(\cdot)$ subject to the power flow equation \cref{eq:powerflow} and additional operational constraints:
\begin{mini!}
	{\substack{\lbrack[\gls{v_bus}]_{\gls{in_bus} \in \gls{bus_set}}, 
			\gls{s_pcc_vec},			
			[\gls{con_load_s_c}]_{\gls{in_con_load} \in \gls{con_load_set}}\rbrack_{\gls{t}\in\gls{t_set}}}}
	{\gls{obj_fun}(\cdot)}
	{\label{eq:opf_problem}}{}
	\addConstraint{\text{\cref{eq:v_pcc}}}{}{\gls{eq:v_pcc} \nonumber}	
	\addConstraint{\text{\cref{eq:s_inj_n}}}{}{\text{Power injections} \nonumber}	
	\addConstraint{\text{\cref{eq:powerflow}}}{}{\text{Power flow equation} \nonumber}
	\addConstraint{|\gls{v_bus_phase}\foft|}{\geq \gls{v_mag_bus_phase_min}}{\!\gls{in_phase}\!\in\!\gls{phase_set_bus},\gls{in_bus} \!\in\! \gls{bus_set_plus}\!,\gls{t} \!\in\! \gls{t_set}\! \label{eq:u_min_bound}}
	\addConstraint{|\gls{v_bus_phase}\foft|}{\leq \gls{v_mag_bus_phase_max}}{\!\gls{in_phase}\!\in\!\gls{phase_set_bus},\gls{in_bus} \!\in\! \gls{bus_set_plus}\!,\gls{t} \!\in\! \gls{t_set}\! \label{eq:u_max_bound}}    	
	\addConstraint{
	\myabs{\sum_{\gls{in_phase} \!\in\! \gls{phase_set}}\gls{s_pcc}^{\gls{in_phase}}\foft}
	}{\leq \gls{s_pcc_sum_mag_max}}{\!\gls{t} \!\in\! \gls{t_set}\label{eq:max_transformer_cap}}
	\addConstraint{ \gls{con_load_p_bus_phase_min} \leq \gls{con_load_p_bus_phase}\foft}{\leq \gls{con_load_p_bus_phase_max}\;}{\gls{in_phase}\!\in\!\gls{phase_set}_{\gls{con_load_bus}},\,\gls{in_con_load} \!\in\! \gls{con_load_set},\,\gls{t} \!\in\! \gls{t_set}\label{eq:con_load_p_bounds}}	
	\addConstraint{\gls{con_load_q_bus_phase_min} \leq \gls{con_load_q_bus_phase}\foft}{\leq \gls{con_load_q_bus_phase_max}}{\gls{in_phase}\!\in\!\gls{phase_set}_{\gls{con_load_bus}},\,\gls{in_con_load} \!\in\! \gls{con_load_set},\,\gls{t} \!\in\! \gls{t_set}\label{eq:con_load_q_bounds}}
\end{mini!}
\Cref{eq:v_pcc,eq:s_inj_n,eq:powerflow} denote the general power network model.
\Cref{eq:u_min_bound,eq:u_max_bound} constrain the voltage magnitude $|\gls{v_bus_phase}\foft|$ to be within a minimal $\gls{v_mag_bus_phase_min} \in \gls{reals}$ and a maximal $\gls{v_mag_bus_phase_max} \in \gls{reals}$ value, according to regulations (e.g., ANSI C84.1). 
\Cref{eq:max_transformer_cap} limits the apparent power injected to the reference bus to be less than $\gls{s_pcc_sum_mag_max} \in \gls{reals_plus}$, typically, to respect the rating of the substation transformer.
\Cref{eq:con_load_p_bounds,eq:con_load_q_bounds} model the characteristics of controllable loads through lower and upper bounds on active power ($\gls{con_load_p_bus_phase_min},\,\gls{con_load_p_bus_phase_max} \in \gls{reals}$), and reactive power ($\gls{con_load_q_bus_phase_min},\,\gls{con_load_q_bus_phase_max} \in \gls{reals}$). The \gls{amod_opf} joint problem described in Section~\ref{subsec:amod_opf_problem} will leverage approximations of the operational constraints in \cref{eq:opf_problem} and include an electricity cost objective term.

This \gls{opf} problem is non-convex because of i) the power flow equation \cref{eq:powerflow} and ii) lower bound constraints on voltage magnitudes \cref{eq:u_min_bound}.
Even the optimization of a balanced single-phase approximation of this problem remains an NP-hard problem \citep{LavaeiLow2012}.

\jvspace{-1em}
\subsection{Convex power flow surrogates} 
\label{sub:power_flow_surrogates}
We desire the joint \gls{amod_opf} problem to be convex and ideally linear to preserve computational tractability.

Hence, we convexify the \gls{opf} problem \cref{eq:opf_problem} using a power flow surrogate that approximates the power flow equation \cref{eq:powerflow} with a convex proxy, making the problem formulation computationally tractable.
Using such a power flow surrogate,
we lose exact knowledge of the indirect variables.

Given the high relevance of the \gls{opf} problem, a vast literature on power flow surrogates exists \citep{MolzahnHiskens2019,Taylor2015}. However, most of these surrogates, as well as comparative studies, consider only balanced single-phase models as typically used in transmission networks \citep{CoffrinGordonEtAl2014}.

For unbalanced three-phase models, only a few power flow surrogates exist, and, to the best of our knowledge, no survey or benchmark classifies the suitability of these surrogates for specific problem structures, such as integration with the \gls{eamod} problem.
To close this gap, we analyzed and compared three promising surrogates.

We compared a convex, \gls{sdp} surrogate \citep{GanLow2014}, the \gls{bfm_sdp}, against two linear surrogates: the \gls{bfm_lp} \citep{GanLow2014,SankurDobbeEtAl2016} and the \gls{lpfm_lp} \citep{BolognaniDoerfler2015}.
We used the charger maximization problem, which maximizes the power delivered to a series of charging stations across a distribution network as a benchmark, as it challenges the surrogates by pushing the network's operational constraints to its limits.
\glsunset{bfm}%
For each surrogate, we evaluated its accuracy in approximating the indirect variables used in  \cref{eq:u_min_bound,eq:u_max_bound,eq:max_transformer_cap}.
Additionally, we  analyzed the resulting constraint violations and computational times
We detail the methodology of our comparison in \citep{Estandia2018} but omit it in this paper due to space limitations.
In summary, \gls{gan_sdp} yielded exact solutions on small instances but performed significantly worse than the other two approaches in both solution quality and computational time for large instances. \gls{bolognani_lp} and \gls{gan_lp} showed a trade-off between solution quality and computational time, with a 91\% reduction on the mean average error in approximating bus voltage magnitudes (\gls{bolognani_lp}) and 97.3\% shorter computational times (\gls{gan_lp}), while neither of both violated the substation rating constraint.
Based on these results, we use the \gls{gan_lp} in this work as it preserves linearity in the joint problem while yielding sufficient solution quality for our mesoscopic study and relatively short computation times.

The \gls{gan_lp} assumes fixed link losses and fixed voltage ratios between phases in a bus \citep{GanLow2014,SankurDobbeEtAl2016}:
let $\gls{i_line_fix} \in \gls{complexes}^{\gls{n_phase_set_line}}$
be the \glsdesc{i_line_fix}.
Let $\gls{v_bus_fix} \in \gls{complexes}^{\gls{n_phase_set_bus}}$ be the \glsdesc{v_bus_fix}.
Then, the \glsdesc{gamma_mat_line}, $\gls{gamma_mat_line} \in \gls{complexes}^{\gls{n_phase_set_line} \times \gls{n_phase_set_line}}$, has entries
\begin{equation*}
\begin{split}
&(\gls{gamma_mat_line}\foft)_{ij} = \frac{((\gls{v_bus_fix}\foft)^{\gls{phase_set_line}})_i}{((\gls{v_bus_fix}\foft)^{\gls{phase_set_line}})_j} \quad i,j \in \arange{\gls{n_phase_set_line}},\\
& \pushright{(\gls{in_line}) \in \gls{line_set},\,\gls{t} \in \gls{t_set}.}
\end{split}
\end{equation*} 
We define the following matrices to ease the notation: 
\begin{align*}
\gls{v_sq_mat}_{\gls{in_bus}}\foft &= \gls{v_bus}\foft \gls{v_bus}\foft\H,\\
\gls{labmda_vec}_{\gls{in_line}}\foft &= \diag{(\gls{v_bus}\foft)^{\gls{phase_set_line}} \gls{i_line}\foft\H}, \\
\gls{ell_mat_fix}_{\gls{in_line}}\foft &= \gls{i_line_fix}\foft\H\gls{i_line_fix}\foft \quad (\gls{in_line}) \in \gls{line_set},\,\gls{t} \in \gls{t_set}.
\end{align*}
Assuming fixed link losses and voltage ratios, the power flow equation \cref{eq:powerflow} admits a linear approximation \citep{ZhaoDallAneseEtAl2017}: 
\begin{align}
\label{eq:power_flow_bfm_1_lp}
&\csum{\gls{in_bus_pre}:(\gls{in_line_pre}) \in \gls{line_set}} \gls{labmda_vec}_{\gls{in_line}}\foft - \diag {\gls{z_mat}_{\gls{in_line_pre}}\gls{ell_mat_fix}_{\gls{in_line}}\foft } -\diag{\gls{v_sq_mat}_{\gls{in_bus}}\foft\gls{y_mat_shunt}_{\gls{in_bus}}\H}\nonumber \\
&  + \gls{s_inj_bus}\foft   =
\csum{\gls{in_bus_post}:(\gls{in_line}) \in \gls{line_set}} (\gls{labmda_vec}_{\gls{in_line}}\foft)^{\gls{phase_set_bus}} \quad \gls{in_bus} \in \gls{bus_set},\,\gls{t} \in \gls{t_set},
\end{align}
\begin{align}
\label{eq:power_flow_bfm_2_lp}
\gls{v_sq_mat}_{\gls{in_bus_post}}\foft = & (\gls{v_sq_mat}_{\gls{in_bus}}\foft)^{\gls{phase_set_line}} - (\gls{gamma_mat}_{\gls{in_line}}\foft \diag{\gls{labmda_vec}_{\gls{in_line}}} \gls{z_mat}_{\gls{in_line}}\H \\
&+ \gls{z_mat}_{\gls{in_line}} (\gls{gamma_mat}_{\gls{in_line}}\foft \diag{\gls{labmda_vec}_{\gls{in_line}}})\H
+ \gls{z_mat}_{\gls{in_line}}\gls{ell_mat_fix}_{\gls{in_line}}\foft\gls{z_mat}_{\gls{in_line}}\H \nonumber \\
& \pushright{(\gls{in_line}) \in \gls{line_set},\,\gls{t} \in \gls{t_set}} \nonumber
\end{align}
The constraints on voltage magnitudes then read:
\begin{align}
\label{eq:voltage_lower_bfm}
    \diag{(\gls{v_sq_mat}_{\gls{in_bus}}\foft)^{\gls{in_phase}}} &\geq (\gls{v_mag_bus_phase_min})^2 \; \gls{in_phase}\in\gls{phase_set_bus},\,\gls{in_bus} \in \gls{bus_set_plus},\,\gls{t} \in \gls{t_set}\\
\label{eq:voltage_upper_bfm}
    \diag{(\gls{v_sq_mat}_{\gls{in_bus}}\foft)^{\gls{in_phase}}} &\leq (\gls{v_mag_bus_phase_max})^2 \; \gls{in_phase}\in\gls{phase_set_bus},\,\gls{in_bus} \in \gls{bus_set_plus},\,\gls{t} \in \gls{t_set}
\end{align}

Now each non-linear term in \cref{eq:opf_problem} can be replaced with a linear approximation to yield the \gls{bfm_lp}: i) the power flow equation \cref{eq:powerflow} with the \gls{bfm} linearization \cref{eq:power_flow_bfm_1_lp,eq:power_flow_bfm_2_lp} and ii) the voltage magnitude constraints \cref{eq:u_min_bound,eq:u_max_bound}  with \cref{eq:voltage_lower_bfm,eq:voltage_upper_bfm}. 
Note that \cref{eq:max_transformer_cap} constrains a complex scalar to lie within a circle of radius \gls{s_pcc_sum_mag_max} in the complex plane.
This constraint is non-linear but convex and can be represented as a second-order cone.
To obtain a \gls{lp}, we approximate the circle with a 12-face regular polygon
\citep{BarnesNagarajanEtAl2017} which covers more than $95\%$ of the circle's area.

The \gls{bfm_lp} has $\gls{n_t}(
\sum_{\gls{in_bus} \in \gls{bus_set}} |\gls{phase_set_bus}|^2  + 
2 \sum_{(\gls{in_line}) \in \gls{line_set}} |\gls{phase_set_line}| +
2 |\gls{phase_set}_{\gls{pcc}}| +
2 \sum_{\gls{in_con_load} \in \gls{con_load_set}} |\gls{phase_set}_{\gls{con_load_bus}}|
)$ decision variables. 
Here, $\gls{n_t}\sum_{\gls{in_bus} \in \gls{bus_set}} |\gls{phase_set_bus}|^2$ is the dominant term since, by assumption, \gls{dist_net} has a tree topology such that $|\gls{bus_set}| = |\gls{line_set}| + 1$. Since voltages are complex-valued (i.e., two components per phase) and $\gls{phase_set_bus}$ has at most three phases, it follows that  $|\gls{phase_set_bus}|^2 \in \gls{bigO}(1)$. 
Thus, the dominant term grows proportional to the number of buses $|\gls{bus_set}|$ and the number of time steps $\gls{n_t}$. In line with this, it admits an upper bound $\gls{bigO}(\gls{n_t}|\gls{bus_set}|)$.

A few comments are in order.
First, we use a linear power flow surrogate which entails the approximation of indirect variables. We discuss its validity and attenuate potential constraint violations in \cref{sec:amod_opf_case_study}.
Second, by using the \gls{gan_lp} surrogate we treat link losses and voltage ratios as fixed parameters.
Previous research has shown that \gls{gan_lp} achieves sufficient accuracy even under the assumption of zero link losses and perfectly balanced voltage ratios \citep{GanLow2014}.
	Our formulation is even more accurate since we use reasonable estimates for the fixed parameters instead of setting them to zero \citep{SankurDobbeEtAl2016}.
\jvspace{-.5em}
\section{Interaction between an electric AMoD system and power distribution networks}
\label{sub:amod_opf}
In this section, we develop a model for the joint optimization of an \gls{eamod} system and a series of \glspl{pdn}. 
Specifically, as an \gls{eamod} system usually spans across multiple (disconnected) \glspl{pdn}, we first introduce the \gls{mopf} problem which combines multiple \gls{opf} problem instances. Then, we formalize the coupling between the \gls{eamod} system and the \glspl{pdn} before we state the joint \gls{amod_opf} problem.
\jvspace{-1em}
\subsection{Multi-\glsentryshort{opf} problem}
\label{sec:multi_opf}
The \gls{mopf} problem couples \gls{n_dist} instances of the \gls{opf} problem and results straightforwardly by extending the constraints for each instance $\gls{in_dist}\in\gls{in_dist_set} = \arange{\gls{n_dist}}$.

We neglect couplings upstream of \gls{pdn} substations through the transmission network as this paper focuses solely on the interaction between an \gls{eamod} system and a series of \glspl{pdn}.
Couplings between the \gls{eamod} system and the power network at the transmission and distribution level occur on very different spatial scales (tens of kilometers vs. hundreds of meters), and result in largely orthogonal effects: specifically, couplings at the transmission level mainly influence bulk electricity prices \cite{RossiIglesiasEtAl2018b}, whereas couplings at the distribution level influence bus voltages and power losses. Accordingly, due to the orthogonal nature of the two couplings, we envision that a nested optimization approach could be used to first address transmission-level couplings through existing algorithms \cite[e.g.,][]{RossiIglesiasEtAl2018b}, and then optimize distribution-level couplings through the tools proposed in this paper.

\jvspace{-1em}
\subsection{Coupling of the electric AMoD system and \glsentrylongpl{pdn}}
\label{sec:amod_opf_coupling}
The charging stations, which appear as controllable loads in the \glspl{pdn}, couple the \gls{eamod} system to the \glspl{pdn} (see \cref{fig:amod_power_interaction}).
Formally, this coupling is established by two functions, \gls{msg} and \gls{map_fun_s_con}, defined below.	

The function $\gls{msg}:\gls{charger_set}\times\gls{t_set} \to \gls{exp_edge_set_charger}$ maps a \glsdesc{in_charger} $\gls{in_charger} \in \gls{charger_set}$ for each time step $\gls{t}\in\gls{t_set}$ to all arcs in \gls{exp_edge_set_charger} that represent charging vehicles at this station:
\begin{equation*}
\gls{msg}(\gls{in_charger},\gls{t}) \!=\! \lbrace (\gls{in_exp_edge}) \!\in\! \gls{exp_edge_set_charger} \!\mid 
\gls{in_exp_node_road} \!=\! \gls{in_exp_node_road_post} \!=\! \gls{in_charger},\,  \gls{in_exp_node_c} \!<\!  \gls{in_exp_node_c_post}, \gls{in_exp_node_t} \!\leq\! \gls{t} \!\leq\! \gls{in_exp_node_t_post} \rbrace.
\end{equation*}
Then, the load at charging station \gls{s} is given by
\begin{equation}
\label{eq:load_at_charger}
\gls{p}_{\gls{in_charger}}\foft = \gls{c_unit}\gls{charge_rate_charger} \,\, \csum{(\gls{in_exp_edge}) \,\, \in \gls{msg}(\gls{in_charger},\gls{t})} \gls{f_0}(\gls{in_exp_edge})\\
\qquad \gls{in_charger}\in \gls{charger_set},\, \gls{t} \in \gls{t_set}
\end{equation}

The function $\gls{map_fun_s_con}:\gls{charger_set} \to (\gls{con_load_set} \times \gls{in_dist_set})$ maps a \glsdesc{in_charger} $\gls{in_charger} \in \gls{charger_set}$ to the associated \glsdesc{in_con_load} $\gls{in_con_load}\in\gls{con_load_set}$ and \glsdesc{in_dist} $\gls{in_dist}\in\gls{in_dist_set}$.
It follows that \glsdesc{in_charger} $\gls{in_charger}$ is attached to bus $\gls{in_bus_map_fun_s_con}$ in \gls{pdn} $\gls{in_dist_map_fun_s_con}$.
As we consider three-phase charging stations, we assume equally distributed loads, that is
\begin{align}
&\gls{con_load_s}^a{}_{,\gls{map_fun_s_con}(\gls{s})}{}\foft = 
\gls{con_load_s}^b{}_{,\gls{map_fun_s_con}(\gls{s})}{}\foft =
\gls{con_load_s}^c{}_{,\gls{map_fun_s_con}(\gls{s})}{}\foft =
\frac{1}{3}\gls{p}_{\gls{in_charger}}\foft \nonumber \\
& \label{eq:con_load_charger} \qquad \qquad \qquad \qquad \qquad \qquad \qquad \qquad  {\gls{in_charger} \in \gls{charger_set},\, \gls{t} \in \gls{t_set}.}
\end{align}
Note that we can model inverters that control the load power factor since $\gls{con_load_q}^{\gls{in_phase}}{}_{,\gls{map_fun_s_con}(\gls{s})}$ must not necessarily be zero. Although the charging station load is distributed equally among phases, loads in distribution networks are inherently unbalanced, which requires an unbalanced distribution model \cite[Ch1.3]{Kersting2002}
Also note that charging stations are commonly modeled to operate at unity power factor (no reactive power consumption) \cite{KeslerKisacikogluEtAl2014}.

\jvspace{-1em}
\subsection{\glsentrytext{amod}-\glsentrytext{opf} problem}
\label{subsec:amod_opf_problem}
The joint \gls{amod_opf} problem results from coupling the \gls{eamod} problem \cref{eq:eamod_problem} with the \gls{mopf} problem
through \cref{eq:load_at_charger,eq:con_load_charger}, namely:
\begin{mini!}
	{\substack{\gls{f_0},[\gls{trip_rate_in_c}]_{\gls{c} \in \gls{c_set}},\\ [\gls{trip_rate_out_t_c}]_{\gls{c} \in \gls{c_set},\gls{t}\in\gls{t_set}}, \gls{dist_vehicle_initial}, \gls{dist_vehicle_final},\\
			\lbrack[\gls{v_bus}]_{\gls{in_bus} \in \gls{bus_set}}, [\gls{i_line_phase}]_{(\gls{in_line}) \in \gls{line_set}},\\
			\gls{s_pcc_vec},
			[\gls{con_load_s_c}]_{\gls{in_con_load} \in \gls{con_load_set}}\rbrack_{\gls{t}\in\gls{t_set},\,\gls{in_dist}\in\gls{in_dist_set}}}}
	{\gls{distance_price}\sum_{(\gls{in_exp_edge}) \in \gls{exp_edge_set}} \gls{d}_{\gls{in_exp_edge_road}} \gls{f_0}(\gls{in_exp_edge})\nonumber}
	{\label{eq:amod_opf_problem}}{}
	\breakObjective{+ \sum_{\gls{t} \in \gls{t_set}} \gls{t_step}
		\sum_{\gls{in_dist} \in \gls{in_dist_set}}
		\gls{electricity_price_dist}\foft
		\sum_{\gls{in_phase} \in \gls{phase_set}} \gls{p}_{\gls{pcc},\gls{in_dist}}^{\gls{in_phase}}\foft \label{eq:amod_opf_problem_obj}}
	\addConstraint{
		\text{\cref{eq:reb_conservation,eq:source_sink_conservation,eq:customer_charge_conservation,eq:road_capacity_adjusted,eq:charger_capacity_physical,eq:dist_vehicle_initial}}\quad
	}{}{\text{\Gls{eamod} system}\nonumber}
	\addConstraint{[\text{\cref{eq:v_pcc,eq:s_inj_n,eq:power_flow_bfm_1_lp,eq:power_flow_bfm_2_lp}}]_{\gls{in_dist}\in\gls{in_dist_set} }\text{ and} \quad}{}{\nonumber}
	\addConstraint{
	[\text{\cref{eq:u_min_bound,eq:u_max_bound,eq:max_transformer_cap,eq:con_load_p_bounds,eq:con_load_q_bounds}}
]_{\gls{in_dist}\in\gls{in_dist_set} \quad }}{}{\text{\glspl{pdn}}\nonumber}
	\addConstraint{\text{\cref{eq:load_at_charger,eq:con_load_charger}}}{}{\text{Coupling from}\nonumber}
	\addConstraint{}{}{\text{charging stations.}\nonumber}
\end{mini!}
The objective \cref{eq:amod_opf_problem_obj} captures operating costs for both the \gls{eamod} fleet and the \glspl{pdn} since we consider full cooperation between both operators. Analogously to the isolated \gls{eamod} problem \cref{eq:eamod_problem_obj}, we consider only rebalancing costs for the \gls{amod} fleet given fixed customer flows.
In each distribution network $d\in\gls{n_dist}$, we account for the electricity cost that results from charging vehicles, uncontrollable loads, and power losses.

Note that our joint problem formulation treats both operators as a single entity, assuming complete information and cooperation.
This assumption is in line with our mesoscopic view and scope to estimate the achievable benefits of coordination and cooperation between the two systems.
We leave the study of game-theoretical aspects to future work where we intend to develop pricing and coordination mechanisms to align the goals of the \gls{eamod} operator and the \gls{pdn} operators, and to leverage distributed optimization algorithms to compute a solution to the \gls{amod_opf} problem \cref{eq:amod_opf_problem} in a distributed manner. Further, our joint model assumes that the \gls{eamod} system is the dominant means of electric transportation, which is in line with our system-level perspective \cite{HannonMcKerracherEtAl2016}. However, the model can readily accommodate other \glspl{ev} by including their traffic flow as residual capacity in \cref{eq:road_capacity_adjusted} and their charging as exogeneous loads in \cref{eq:opf_problem}.

\jvspace{-.5em}
\section{Case study in Orange County, CA}
\label{sec:amod_opf_case_study}
We evaluate the impact of an \gls{eamod} system on the \glspl{pdn} and the benefit of optimized joint coordination through a case study in \gls{oc}.
Our case study considers commuting trips within the cities of Fountain Valley, Irvine, North Tustin, Orange, Santa Ana, Tustin, and Villa Park. 
In the following, we detail our data~(\ref{sec:case_study_parameters}), outline the experimental design (\ref{sec:amod_opf_experimental_design}) and, finally, discuss our results (\ref{sec:amod_opf_results}).
\jvspace{-5mm}
\begin{figure}[!hb]
	\centering
	\includegraphics[width=0.55\linewidth]{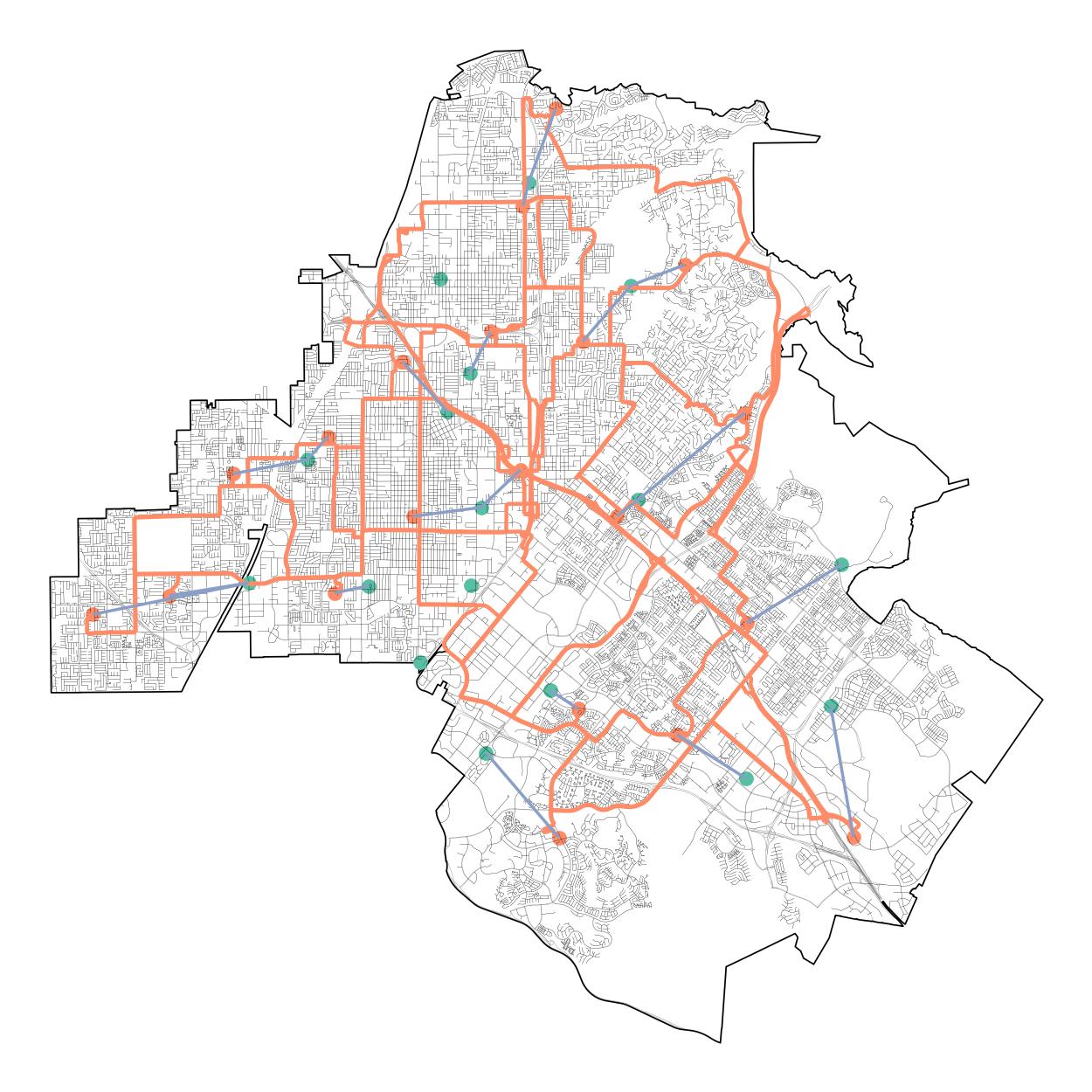}
	\caption{\label{fig:orange_county} Area considered in the \gls{oc} case study. 
		The aggregated road network is shown in orange, representing vertices as dots and arcs as lines. 
		Green dots show the substation locations. 
		Blue lines show the assignment of a charging station to its closest substation.}
\end{figure}
\jvspace{-2mm}

\jvspace{-1em}
\subsection{Model parameters}
\label{sec:case_study_parameters}
We focus on an eight-hour commuting cycle from $\unit[5]{am}$ to $\unit[1]{pm}$ on July 3, 2015 discretized into six-minute time steps, such that $|\gls{t_set}| = 80$. As we do not consider future grid storage devices, which would charge/discharge over the span of a day, an eight-hour horizon is sufficient to model the power system. We chose the time discretization to be close to the traversal time of the shortest road link. As the power system considers hourly prices and excludes transient effects, six-minute time steps are more than sufficient to model \glspl{pdn} for a mesoscopic analysis.
For this period, we model the charging station and transportation networks at a mesoscopic aggregation level that allows a sufficient level of detail to analyze the interaction between an \gls{eamod} system and the \glspl{pdn}, and ensures computational tractability.

\paragraph{Transportation network data} we derive trip demand from Census Tract Flow data from the 2006-2010 American Community Survey.
From these data, we take the estimated commuting flows between the $143$ census tracts that are part of our case study.
To align the granularity of aggregated charging station network representations and census tracts, we cluster the 143 census tracts into 20 larger areas using a k-means algorithm.
We neglect commuting flows if they start or end outside the area of our case study or if they start and end within the same cluster since these types of flows cannot be accurately represented in our model.
Our planning horizon comprises $\numprint{122219}$trips (32.8\% of the total daily trips).

The problem of fleet sizing for (electric) \gls{amod} systems \cite{VazifehSantiEtAl2018} is beyond the scope of this paper. 
For this case study, we heuristically selected a sufficient fleet size, large enough to keep the \gls{amod_opf} problem \cref{eq:amod_opf_problem} feasible with only a small number of idle vehicles and corresponding to 140\% of the peak concurrent number of passenger-carrying trips.

We create an aggregated road network based on OpenStreetMap data with the same granularity as the trip demand data.
For this, we select  the road network vertices closest to the centroids of the census tract clusters and add arcs between those vertices if a connection exists in the real road network.
We obtain an aggregated road network with 20 vertices and 
$\numprint{76}$arcs (see \cref{fig:orange_county}), which captures vehicle travel and charging between the separate \glspl{pdn} of the case study region.
Note that computational complexity limits our model to coarse road networks; this is discussed in detail below.
For each aggregated road network vertex, we consider three-phase 50-kilowatt DC fast charging stations with $\gls{charger_cap} = 40$ plugs in total. 
Accordingly, each vertex has a charging station with a maximum load of two megawatts ($0.66$ megawatts per phase). 

\paragraph{Electric vehicle data} we consider a homogeneous vehicle fleet based on the characteristics of the 2018 Nissan Leaf which has a $40$-kilowatt-hour battery and a range of $240$ kilometers.
Based on fast-charging guidelines, we reduce a vehicle's battery capacity and its range to $80$ percent of their original values \citep{ChenKockelmanEtAl2016}, and discretize this effective battery capacity into $\gls{n_c} = 40$ levels, resulting in energy discretizations of 0.8 kWh which remains close to the energy necessary to traverse the lowest energy road link. 
To account for the possibility that vehicles might not start the day with fully charged batteries, we set the \gls{soc} at $\gls{t} = 1$ to $50\%$.
Furthermore, we require vehicles to recharge the amount of energy used over a planning horizon such that the final \gls{soc} must be at a minimum 50\% again.
We set the \glsdesc{distance_price} to $\gls{distance_price} = \unitfrac[0.3]{USD}{km}$ \citep{BTS2016}.

\paragraph{Power distribution networks data} we use a \gls{gld} model of the PL-1 distribution network, a primary feeder operated by the \gls{pge} available for research purposes \citep{PGEC}, as a proxy for (sub-)urban distribution networks.
The network comprises 322 buses and operates at a nominal voltage of $12.6$ kilovolts.
We set the uncontrollable loads to the model's time-varying loads.

We take the location of substations from the utility's data \citep{SCE} and attach a model of the PL-1 distribution network to each substation.
We set the electricity price at each substation to the corresponding locational marginal price \citep{CISO} and conservatively assume a base load utilization of 75 percent at the substation transformer. 
Typically, distribution networks are operated at $50$ to $75$ percent of their load capacity so that loads can be transferred from one distribution network to another if needed \cite{SchneiderChenEtAl2008}.
Accordingly, we set the \glsdesc{s_pcc_sum_mag_max} \gls{s_pcc_sum_mag_max} to $1/0.75$ times the value of the peak base load (i.e., without charging stations), yielding $\gls{s_pcc_sum_mag_max} = \unit[10.42]{MVA}$.
In addition, we set the lower voltage magnitude limit to $0.96$ per-unit and the upper limit to $1.04$ per-unit, which is $0.01$ per-unit tighter than required by ANSI C84.1 to allow for the voltage drop in the secondaries of the network.

We connect each charging station to the distribution network whose substation is nearest.
Since no data on the coordinates of the distribution network buses exist, we randomly attach the charging station to one of the \gls{pdn} buses.
Thus, the \gls{pdn} is the same for each substation, except for the varying number and location of charging stations.
In total, we consider $14$ distribution networks, each with one or two charging stations.

We set the price of electricity at each charging station to be equal to the electricity price at the respective substation, such that $\gls{price}_{\textrm{el},\gls{in_charger}}\foft =  \gls{price}_{\textrm{el},\gls{in_dist}_{\gls{map_fun_s_con}(\gls{s})}}\foft$ holds. 
Since we focus on the total benefit from a system perspective and treat both operators as a single entity, only the spatial variation of electricity prices that are closely linked to the substation prices affects our solution. 

Some comments on the distribution network modeling are in order. First, we used the same network model and load values for each distribution network, considering loads from a single summer day. 
As \glspl{pdn} are treated as critical infrastructure and load data is usually confidential to protect customers, more accurate data is not publicly available for research purposes \citep{MarcosDomingoEtAl2017}. 
However, our model can be rerun with more accurate data at any time.
Second, we set the electricity price at each substation to the corresponding locational marginal price. 
Locational marginal prices result from the power consumption at the transmission grid level. As our focus is on the interaction of the \gls{eamod} fleet with the distribution grids and the power used for recharging represents only a negligible fraction at the transmission grid level, neglecting the impact of this consumption on the marginal prices only minimally affects the accuracy of our results.
Third, we assume the electricity price for charging at a certain station to be equal to the electricity price at the respective substation.
Neglecting the possible difference in electricity prices among nodes in a single distribution network is consistent with our mesoscopic transportation model.

The resulting \gls{amod_opf} problem has \numprint{6224240} decision variables, \numprint{1463600} from the \gls{eamod} part and \numprint{4760640} from the \gls{mopf} part.
Since the \gls{mopf} part comprises \gls{n_dist} \glspl{pdn}, the number of variables in it admits the upper bound $\gls{bigO}(\gls{n_t}\gls{n_dist}|\gls{bus_set}|)$.
Thus, the number of decision variables in the whole \gls{amod_opf} problem admits the following upper bound: $\gls{bigO}(\gls{n_t}(\gls{n_c}|\gls{road_node_set}|^2 +\gls{n_dist}|\gls{bus_set}|)).$
Recall that the complexity of solving the LP with an interior point method is polynomial in the number of variables with an exponent lower than 3.5 (depending on the implementation) \cite{Gondzio2012}.
Nominally, the size of the \gls{eamod} part of the problem increases quadratically with the number of road vertices. 
However, if more vertices are added for the same area, the road segment arcs will become shorter, and \gls{n_t} and \gls{n_c} should be increased to capture the reduced travel duration and energy consumption in the shorter road segment arcs.
Thus, in practice, the \gls{eamod} part of the problem grows more than quadratically with the number of road vertices.
This limits our formulation to coarse road networks.
In future work, we will explore methods that improve the scalability of the \gls{amod_opf} problem, extending its applicability to finer networks.

\jvspace{-1em}
\subsection{Experimental design}
\label{sec:amod_opf_experimental_design}
To quantify the impact of an \gls{eamod} system on the \glspl{pdn} and the benefit of optimized joint coordination, our experiments consider two cases. First, we analyze the impact of an \gls{eamod} system on the \glspl{pdn} without coordination, i.e., the uncoordinated case.
This study shows how \gls{eamod} systems can negatively affect \glspl{pdn}.
Then, we focus on the coordinated case in which the \gls{eamod} system and the distribution networks are jointly optimized.
Comparing the results of both cases allow us to quantify the potential of optimized coordination between these systems. 
In both cases we generate results as follows:
\paragraph{Computing controllable loads} we determine the load at each charging station that results from the operation of the \gls{eamod} system.
Depending on the studied case, we solve either \cref{eq:eamod_problem} (uncoordinated) or \cref{eq:amod_opf_problem} (coordinated).
\paragraph{Solving the power flow equation} to assess the quality of a solution from step (a), we solve the exact power flow equation \cref{eq:powerflow} to derive the true values of the indirect variables (i.e., complex power injection at the reference bus and complex voltage in all other buses).
\paragraph{Evaluating constraint violations} in step (a), we determine controllable loads without an exact model of the \glspl{pdn} as it is either neglected (uncoordinated case) or approximated (coordinated case).
Hence, it is often the case that solutions do violate some of the constraints.
To quantify these violations, we evaluate integral constraint violations as we consider a time-variant model. Specifically, regulations require voltage magnitudes to be kept within a given percentage of a nominal value (e.g., ANSI C84.1).
Hence, we analyze the \glsdesc{u_viol}
\begin{equation*}
\gls{u_viol} = \gls{t_step}\sum_{\gls{t} \in \gls{t_set}} \sum_{\gls{in_dist} \in \gls{in_dist_set}} \sum_{\gls{in_bus} \in \gls{bus_set_plus}_{\gls{in_dist}}} \sum_{\gls{in_phase} \in \gls{phase_set_bus_d}} 
|\gls{u_viol_bus_phase_d}\foft|
\end{equation*}
where 
\begin{equation*}
\begin{split}
\gls{u_viol_bus_phase_d}\foft &= \min(\gls{u_bus_phase_d}\foft - \gls{v_mag_bus_phase_min_d},\, 0) \\
	 &+ \max(\gls{u_bus_phase_d}\foft - \gls{v_mag_bus_phase_max_d},\, 0)
\end{split}
\end{equation*}
is the \glsdesc{u_viol_bus_phase_d}. 
Note that \gls{u_viol_bus_phase_d} is negative when the voltage magnitude is lower than \gls{v_mag_bus_phase_min_d}, positive when it is larger than \gls{v_mag_bus_phase_max_d}, and zero when it is in-between.
Additionally, substations typically connect distribution networks to the higher-voltage transmission network, requiring a transformer to lower the voltage.
To avoid overloading this transformer, the power draw must be less than the transformer rating. Hence, we analyze the \glsdesc{s_pcc_sum_mag_max_viol} 
\begin{equation*}
\gls{s_pcc_sum_mag_max_viol} = \sum_{\gls{t} \in \gls{t_set}} \gls{t_step} \,
\sum_{\gls{in_dist} \in \gls{in_dist_set}} \gls{s_pcc_sum_mag_max_viol_d}\foft
\end{equation*}
where
\begin{equation*}
\gls{s_pcc_sum_mag_max_viol_d}\foft = \max(|\sum_{\gls{in_phase} \in \gls{phase_set}_{0,\gls{in_dist}}} \gls{s_pcc_d}^{\gls{in_phase}}\foft| - \gls{s_pcc_sum_mag_max_d},\, 0)
\end{equation*}
is the \glsdesc{s_pcc_sum_mag_max_viol_d}.
\paragraph{Evaluating energy consumption and cost} we analyze the energy consumption of the \gls{eamod} system and its cost.
The \glsdesc{E_total} \gls{E_total}, which includes the energy consumed by exogenous loads and the \gls{eamod} system, results from summing the energy draw of all substations. The \glsdesc{E_total_base} \gls{E_total_base} results analogously without considering an \gls{eamod} system. Consequently, the difference of \gls{E_total} and \gls{E_total_base} represents the \glsdesc{E_amod}: 
\begin{align*}
\gls{E_amod} &= \gls{E_total} - \gls{E_total_base} \\
&= \sum_{\gls{t} \in \gls{t_set}} \gls{t_step} \, 
\sum_{\gls{in_dist} \in \gls{in_dist_set}}
\sum_{\gls{in_phase} \in \gls{phase_set}_{0,\gls{in_dist}}}
(\gls{p_pcc_d}^{\gls{in_phase}}\foft - 
\gls{p}_{\gls{base},\gls{in_dist}}^{\gls{in_phase}}\foft).
\end{align*}
Here, $\gls{p}_{\gls{base},\gls{in_dist}}^{\gls{in_phase}} \in \gls{reals}$ is the power drawn in phase $\gls{in_phase} \in \gls{phase_set}_{\gls{pcc}}$ from substation $\gls{in_dist} \in \gls{in_dist_set}$ in the base case.

Due to losses in the distribution networks, not all of \gls{E_amod} relates to charging stations.
The \glsdesc{E_amod_charge} is given by
\begin{equation*}
\gls{E_amod_charge} = \sum_{\gls{t} \in \gls{t_set}} \gls{t_step} \,
\sum_{\gls{in_dist} \in \gls{in_dist_set}}
\sum_{\gls{in_con_load} \in \gls{con_load_set_d}} \sum_{\gls{in_phase} \in \gls{phase_set}_{\gls{con_load_bus},\gls{in_dist}}} \gls{con_load_p_bus_phase_d}\foft.
\end{equation*}	
The difference between \gls{E_amod} and \gls{E_amod_charge} represents the \glsdesc{E_amod_losses}:
\begin{equation*}
\gls{E_amod_losses} = \gls{E_amod} - \gls{E_amod_charge}.
\end{equation*}
Analogously, the cost of these losses is given by 
\begin{equation*}
\gls{electricity_cost_losses} = \gls{electricity_cost_amod} - \gls{electricity_cost_charging}
\end{equation*}
where \gls{electricity_cost_amod} is given by
\begin{equation*}
\gls{electricity_cost_amod} = \sum_{\gls{t} \in \gls{t_set}} \gls{t_step}
\sum_{\gls{in_dist} \in \gls{in_dist_set}}
\gls{price}_{\textrm{el},\gls{in_dist}}\foft
\sum_{\gls{in_phase} \in \gls{phase_set}_{0,\gls{in_dist}}} 
(\gls{p_pcc_d}^{\gls{in_phase}}\foft - \gls{p}_{\gls{base},\gls{in_dist}}^{\gls{in_phase}}\foft)
\end{equation*}
and \gls{electricity_cost_charging} is the cost of \gls{E_amod_charge}:
\begin{equation*}
\begin{split}
&\gls{electricity_cost_charging} = 
\sum_{\gls{t} \in \gls{t_set}} \gls{t_step} \,
\sum_{\gls{in_dist}  \in \gls{in_dist_set}}
\gls{price}_{\textrm{el},\gls{in_dist}}\foft
\sum_{\gls{in_con_load} \in \gls{con_load_set_d}} \sum_{\gls{in_phase} \in \gls{phase_set}_{\gls{con_load_bus},\gls{in_dist}}} \gls{con_load_p_bus_phase_d}\foft.
\end{split}
\end{equation*}	

Our implementation builds on top of the authors' AMoD Toolkit\footnote{\url{https://github.com/StanfordASL/AMoD-toolkit}} which relies on \gls{yalmip} \citep{Loefberg2004} to formulate and solve \gls{eamod} problems. Additionally, we built a general codebase for unbalanced \gls{opf} problems, the \gls{uot}\footnote{\url{https://github.com/StanfordASL/unbalanced-opf-toolkit}}.
To support future research in this field, we released both the AMoD Toolkit and the \gls{uot} under an open-source license. 
\jvspace{-1em}
\subsection{Results and discussion}
\label{sec:amod_opf_results}
\begin{figure}
	\centering
	\begin{subfigure}[b]{0.45\linewidth}
		\centering
		\includegraphics[width=\textwidth]{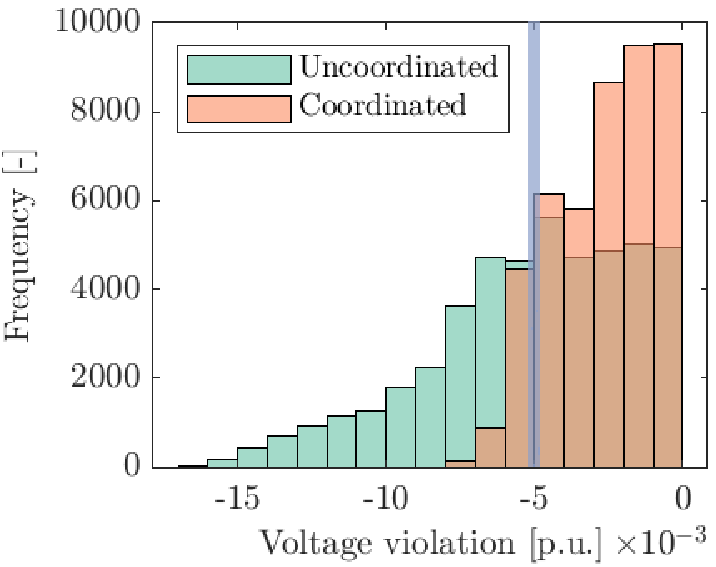}
		\caption{\label{fig:voltage_violations} Voltage magnitude}
	\end{subfigure}
	\begin{subfigure}[b]{0.45\linewidth}
		\centering
		\includegraphics[width=\textwidth]{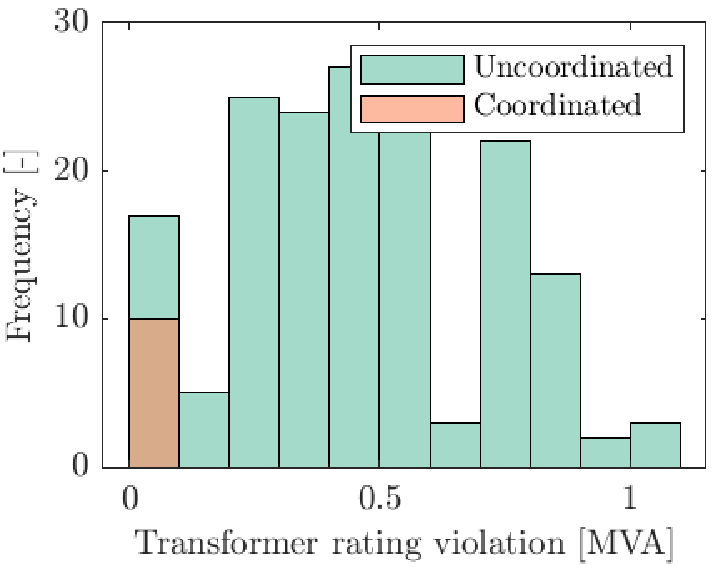}
		\caption{\label{fig:substation_capacity_violations} Substation transformer rating}
	\end{subfigure}	
	\caption{\label{fig:constraint_violations} Histograms of voltage magnitude (\gls{u_viol_bus_phase_d}) and substation transformer rating (\gls{s_pcc_sum_mag_max_viol_d}) violations.
	For clarity, we do not show the cases where the violation is zero.
	All voltage magnitude violations are negative because the upper limit \gls{v_mag_bus_phase_max_d} is never exceeded.
	The vertical line indicates the threshold for serious voltage magnitude violation events (i.e., those exceeding $0.005$ p.u.)
	For both quantities, constraint violations are significantly lower in the coordinated case.
	}
\end{figure}

Following our experimental design, we evaluate constraint violations (\cref{fig:constraint_violations}), as well as energy consumption and costs. 
\Cref{table:summary} summarizes the key results.

\Cref{fig:voltage_violations} shows a histogram with all voltage magnitude constraint violations \gls{u_viol_bus_phase_d}; each event represents the constraint being violated in one phase during one of the six-minute time steps.
The base case shows no violations and, hence, is not plotted. 
In contrast, violations appear in both cases that include the \gls{eamod} system.
ANSI C84.1, the power quality standard for voltage ranges used across the United States, advises that service voltage violations must be limited in extent, frequency, and duration.
Optimized coordination between the \gls{eamod} system and the \glspl{pdn} helps to decrease voltage constraint violations significantly. 
The number of voltage constraint violations is reduced by $\numprint{3.85}$percent in the coordinated case, from $\numprint{46910}$ to $\numprint{45106}$\unskip.
Notably, coordination reduces the number of serious violation events (i.e., those exceeding $0.005$ p.u. which are the most concerning, see \cref{fig:voltage_violations}) by $\numprint{74.85}$percent, from $\numprint{21734}$to $\numprint{5467}$\unskip.
All in all, there is a $\numprint{50.28}$percent reduction in \glsdesc{u_viol}, from $\numprint{24.04}$per-unit hour to $\numprint{11.95}$per-unit hour.
Consequently, coordination between the two systems helps to achieve better compliance with regulations that require the voltage magnitude to be kept close to its nominal value.

\Cref{fig:substation_capacity_violations} shows a histogram with all substation transformer rating violations \gls{s_pcc_sum_mag_max_viol_d}; each event represents the constraint being violated in one substation transformer during one of the six-minute time steps.
Optimized coordination nearly eliminates substation capacity constraint violations, reducing their count by $\numprint{94.05}$percent from $\numprint{168}$to $\numprint{10}$\unskip.
The number of substations that experience a transformer rating violation is reduced from six to two.
All in all, there is a $\numprint{99.71}$percent reduction in \glsdesc{s_pcc_sum_mag_max_viol}, from $\numprint{7.89}$mega volt-ampere hour to $\numprint{0.02}$mega volt-ampere hour.

Transformers represent a significant investment by utilities.
For example, installing a transformer with a rating similar to the one used in this case study ($\gls{s_pcc_sum_mag_max} = \unit[10.42]{MVA}$) has a cost in the order of 1.7 million USD \cite{Sarker2017}.
Given transformers' substantial cost, increasing their useful life by reducing transformer capacity threshold violations (as done by coordination) can lead to significant monetary savings for utilities.
We leave the precise quantification of these savings for future research.

\begin{figure}
\jvspace{.3em}
	\centering
	\includegraphics[width=0.55\linewidth]{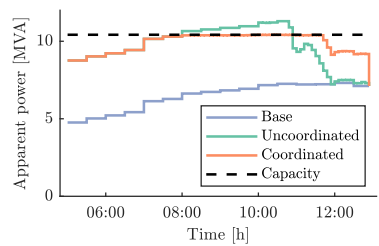}
	\caption{\label{fig:substation_load_bryan} Load at one representative substation and the corresponding transformer rating. 
	The base case shows the load from uncontrollable loads. 
	The uncoordinated and coordinated cases show increased load due to charging vehicles.
	The substation transformer rating is exceeded in the uncoordinated case.
	In the coordinated case, charging vehicles later during the day resolves the violation.}
\end{figure}

\Cref{fig:substation_load_bryan} shows the load at one representative substation along with the applicable transformer rating.
The load is shown for the three cases: base, uncoordinated and coordinated.
The base case represents the substation load arising from the uncontrollable loads.
The other two cases show higher loads due to the recharging vehicles.
In the uncoordinated case, there is a significant transformer rating violation between 8 am and 11 am.
Coordination helps to resolve the violation, as charging loads that exceed the capacity constraint are shifted to later time steps.	

\begin{figure}
	\centering
	\includegraphics[width=0.55\linewidth]{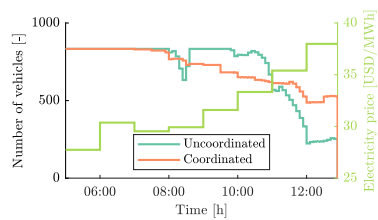}
	\caption{\label{fig:charging_vehicles} Number of charging vehicles and electricity price over time.
	Coordination shifts the charging load to later time steps when the electricity price is higher.
	The resulting cost increase is the price paid for reducing system constraint violations, which improves voltage profiles and prolongs transformer life.
	}
\end{figure}

\Cref{fig:charging_vehicles} shows the number of charging vehicles and the electricity price over time.
The coordinated case shows steady charging activity after 11 am.
In contrast, charging activities decrease significantly after 11 am in the uncoordinated case.
The charging activity mirrors the substation load in \cref{fig:substation_load_bryan} which is higher for the coordinated case in later time steps.
The increased charging activity later in the day and the ensuing load leads to increased electricity expenditure as the electricity price is higher later in the day.

\cref{table:summary} shows the impact of coordinating an \gls{eamod} fleet with \glspl{pdn}.
The total operational costs of the \gls{eamod} system during the studied 8-hour time span increase slightly by 
$\numprint{3.13}$percent ($\numprint{3329.61}$USD).
Rebalancing costs show an increase of $\numprint{3.28}$percent ($\numprint{3206.47}$USD) as vehicles charge at more distant charging stations due to an increase in rebalancing detours. 
The shift of charging activity to later in the day due to coordination causes electricity costs to increase by $\numprint{1.42}$percent ($\numprint{123.15}$USD). 
The small increase in operational costs reflects the price paid for reducing system constraint violations, which improves voltage profiles and prolongs transformer life.

The energy delivered to the charging stations (see \cref{table:summary}) increases by $\numprint{1.82}$megawatt-hour ($\numprint{0.68}$percent) in the coordinated case because of increased rebalancing detours.
However, the energy attributable to the \gls{eamod} system consumed at the substations increases only by $\numprint{1.24}$megawatt-hour ($\numprint{0.44}$percent).
The difference of $\numprint{0.58}$megawatt-hour is due to energy losses being reduced by $\numprint{5.24}$percent.
Reduced energy losses reflect more efficient power distribution: a greater share of the energy leaving the substations reaches the charging stations in the coordinated case ($\numprint{96.29}$percent compared with $\numprint{96.07}$percent).

The optimization was performed on an AWS r4.xlarge instance (4 vCPU at $\unit[2.3]{GHz}$, $\unit[30.5]{GB}$ RAM). The \gls{amod_opf} problem \cref{eq:amod_opf_problem} was solved in 554 iterations over $8.1$hours, whereas solving the \gls{eamod} problem \cref{eq:eamod_problem} took 51 iterations over $0.7$hours using Gurobi Optimizer.
Thus, the presented solution approach is currently not suitable for real-time operations--the design of an operational version of this framework is left for future research. 
One potential avenue for reducing the computation time would be improving the scaling of the \gls{amod_opf} problem \cref{eq:amod_opf_problem} to reduce the number of iterations required by the solver. 
Despite the computation times, the mesoscopic analyses presented herein can be used to identify bottlenecks in \glspl{pdn} that point at necessary grid extension investments. 
Additionally, a grid operator can use this approach to compute the amount of spinning reserves needed to hedge on the day-ahead market to secure a reliable operation of its \glspl{pdn}.

\begin{table}
\centering
{\small
\begin{tabular}{llccc}
\hline
 & Unit & Uncoord & Coord & Change\!\!\\
\hline
Voltage violation & p.u. h & $\num{24.04}$ & $\num{11.95}$ & $\num{-50.28}$\% \\
Capacity violation & MVAh & $\num{7.89}$ & $\num{0.02}$ & $\num{-99.71}$\% \\
\hline
Electricity cost, & USD & $\num{8.35}$k & $\num{8.49}$k & $\num{1.67}$\% \\
charging &&&&\\
Electricity cost, & USD & $\num{0.35}$k & $\num{0.33}$k & $\num{-4.59}$\% \\
losses &&&&\\
Electricity cost, & USD & $\num{8.69}$k & $\num{8.82}$k & $\num{1.42}$\% \\
AMoD &&&&\\
Rebalancing cost  & USD & $\num{97.79}$k & $\num{101.00}$k & $\num{3.28}$\% \\
Total cost, AMoD & USD & $\num{106.49}$k & $\num{109.82}$k & $\num{3.13}$\% \\
\hline
\gls{charger_energy} & MWh & $\num{268.82}$ & $\num{270.63}$ & $\num{0.68}$\% \\
\gls{losses_energy} & MWh & $\num{11.01}$ & $\num{10.43}$ & $\num{-5.24}$\% \\
\gls{eamod_energy} & MWh & $\num{279.82}$ & $\num{281.06}$ & $\num{0.44}$\% \\
\hline
\end{tabular}
}
\caption{Impact of coordinating an \gls{eamod} fleet with \glspl{pdn}. 
Coordination significantly reduces constraint violations at the cost of slightly higher operational costs.}
\label{table:summary}
\end{table}
\jvspace{-.5em}
\section{Conclusion}
\label{sec:conclusion}
We presented the \gls{amod_opf} problem, which integrates an \gls{eamod} problem with a \gls{mopf} problem.
In this context, we discussed power flow surrogates to obtain a computationally tractable convex problem formulation. 
The resulting \gls{amod_opf} problem allows one to assess the achievable benefit of coordinating an \gls{eamod} system and a series of \glspl{pdn}.
With this methodological framework, we investigated the impact of an \gls{eamod} system on the \glspl{pdn}. 
Herein, we especially focused on the benefits of coordination between  the two systems and discussed results for a case study in \gls{oc}.
We showed that in an uncoordinated system, the \gls{eamod} fleet negatively affects the distribution networks: the charging behavior of the \gls{eamod} vehicles caused overloads at substation transformers and violated (lower) voltage magnitude limits.
Furthermore, we showed that a coordinated system helps to balance the load in the \glspl{pdn} in time and space.
Specifically, link losses were slightly reduced, substation overloads were nearly eliminated, and voltage violations were halved. 
Nonetheless, these reductions in constraint violations increased the cost of operating the \gls{eamod} system by $\numprint{3.13}$percent caused by vehicles driving to charge in less congested but more distant stations and charging when electricity prices are higher.
This indicates that distribution networks can support more electric vehicles before upgrades are needed if the vehicles are charged in coordination with exogenous loads in the \glspl{pdn}. Due to our system-optimal objective, these findings remain an assessment of the overall benefit of coordination between an \gls{eamod} fleet and \glspl{pdn}.

Our findings open the field for multiple directions of future research. 
First, our \gls{amod_opf} problem is mesoscopic and assumes perfect knowledge of future loads and trip requests. To design a real-time algorithm, the integration of forecasts to capture the stochastic nature of the problem is an interesting avenue for further research.
Second, we modeled the operators of the \gls{amod} fleet and the \glspl{pdn} as a single entity, implying full cooperation. In future work, one should address the interplay between these two stakeholders, with the goal of designing incentive mechanisms, and investigate market dynamics, e.g., the price of stability and the price of anarchy.
Third, our case study provides preliminary results about the benefit of coordinating \gls{eamod} fleets with \glspl{pdn}. To provide  decision support to practitioners, additional case studies that capture different \glspl{pdn}, different road network characteristics, varying instance sizes, and distributed renewable energy generation are required.
Fourth, our case study did not consider the \glspl{ev} potential to feed power back into the \gls{pdn}. Hence, extending our modeling approach for vehicle-to-grid options, evaluating regulation and operating reserve potentials, remains a promising avenue for future research.

\jvspace{-.5em}
\section*{Acknowledgments}
The authors thank  Saverio Bolognani, David Chassin and Raffi Sevlian for sharing their insights on power systems. 
A. Estandia was supported by the Zeno Karl Schindler Foundation with a Master's Thesis Grant. 
This research was supported by the National Science Foundation under the CAREER and CPS programs, the Toyota Research Institute (TRI), and the Stanford Bits \& Watts EV50 Project. This article solely reflects the opinions and conclusions of its authors and not NSF, TRI, any other Toyota entity, or Stanford. 

\begin{appendices}
\crefalias{section}{appsec}
\jvspace{-1em}
\section{Nomenclature}
\label{app:nomenclature}
{\small
\printglossary[type=amod]
\jvspace{-1em}
\printglossary[type=opf]
}	
\end{appendices}
\jvspace{-2mm}\textbf{}


%
\jvspace{-.5em}
{\small
\printbibliography
}

\begin{IEEEbiography}[{\includegraphics[width=1in,height=1.25in,clip,keepaspectratio]{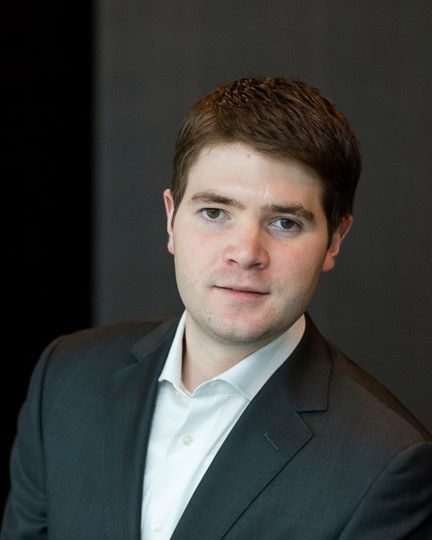}}]{Alvaro Estandia} is a Software Engineer at Marain Inc. 
He earned an MSc in Robotics in 2018 and a BSc in Mechanical Engineering in 2015, both from ETH Zurich.
He develops software to simulate and algorithms to control fleets of electrical autonomous vehicles providing mobility-on-demand in urban environments.
More broadly, he is interested in the applications of optimization for improving the  performance of transportation systems and the power network.
\end{IEEEbiography}
\jvspace{-3em}
\begin{IEEEbiography}[{\includegraphics[width=1in,height=1.25in,clip,keepaspectratio]{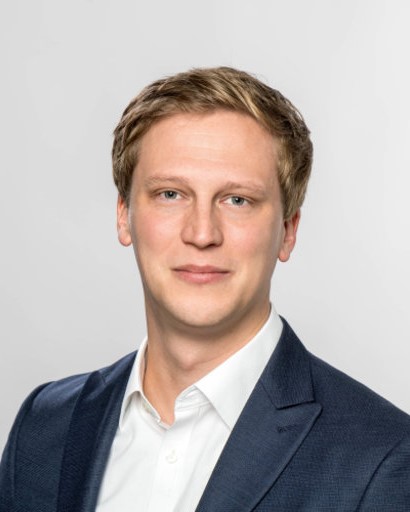}}]
{Maximilian Schiffer} is an Assistant Professor of Operations and Supply Chain Management at Technical University of Munich. He received a Ph.D. degree in Operations Research from RWTH Aachen University in 2017. His main research interests are in operations research, machine learning, and intelligent systems, with an emphasis on transportation and logistics topics, especially electric vehicles and autonomous systems. He is a recipient of the INFORMS TSL Dissertation Prize and the GOR Doctoral Dissertation Prize.
\end{IEEEbiography}
\jvspace{-3em}
\begin{IEEEbiography}[{\includegraphics[width=1in,height=1.25in,clip,keepaspectratio]{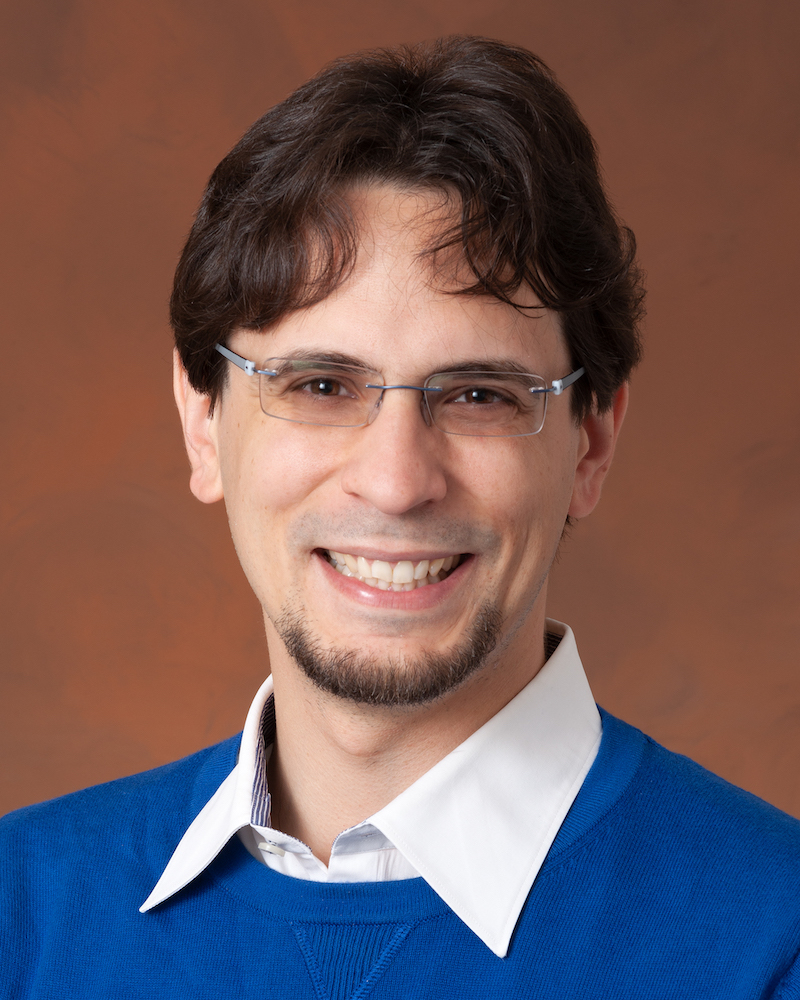}}]{Federico Rossi} is a Robotics Technologist at the Jet Propulsion Laboratory, California Institute of Technology.
He earned a Ph.D. in Aeronautics and Astronautics from Stanford University in 2018, a M.Sc. in Space Engineering from Politecnico di Milano, and the Diploma from the Alta Scuola Politecnica in 2013.
His research focuses on optimal control and distributed decision-making in multi-agent robotic systems, with applications to planetary exploration and coordination of fleets of self-driving vehicles for autonomous mobility-on-demand.
\end{IEEEbiography}
\jvspace{-3em}
\begin{IEEEbiography}[{\includegraphics[width=1in,height=1.25in,clip,keepaspectratio]{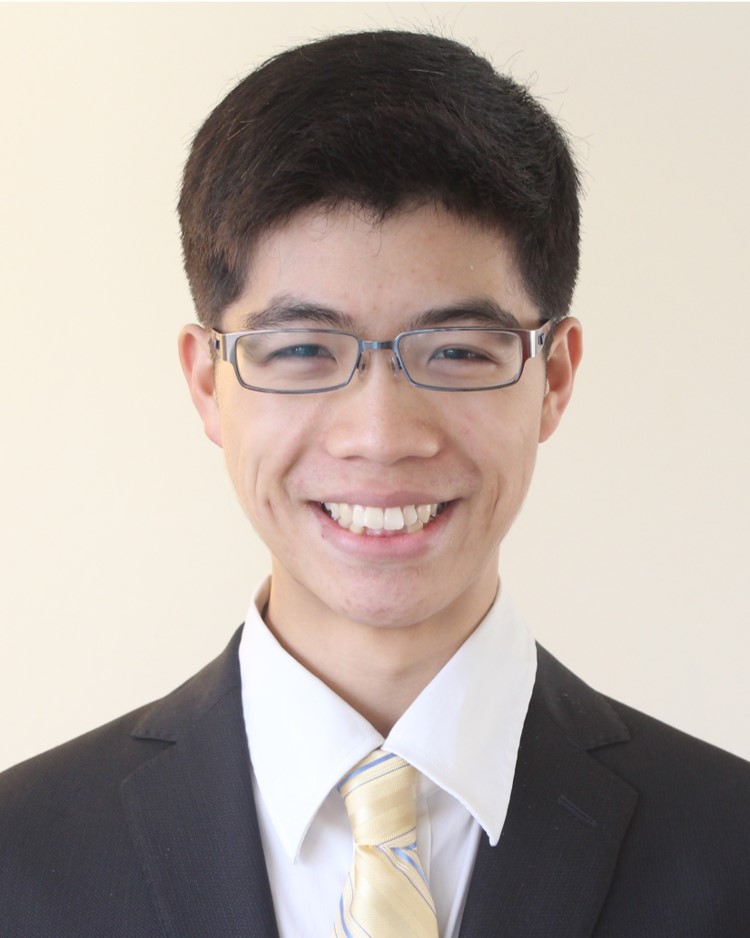}}]{Justin Luke} is a Ph.D. Candidate at Stanford University in the Autonomous Systems Laboratory and Sustainable Systems Laboratory. He earned a B.S. in Energy Engineering from the University of California, Berkeley in 2018. His research focuses on optimization methods for integration of electric autonomous mobility-on-demand fleets into the electricity grid, particularly in scenarios with high-penetration of renewable generation.
\end{IEEEbiography}
\jvspace{-3em}
\begin{IEEEbiography}[{\includegraphics[width=1in,height=1.25in,clip,keepaspectratio]{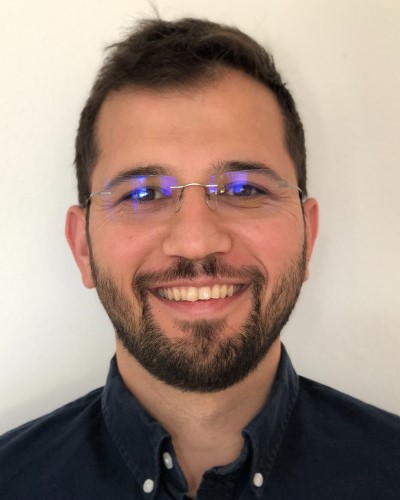}}]{Emre Kara} received the Ph.D. degree from the Carnegie Mellon University focusing on infrastructure systems, machine learning, and data science. He is currently leading the engineering and data science efforts with eIQ Mobility. His research interests include data-driven methods to integrate HVAC, electric vehicles, and battery storage systems into the electricity grid as flexibility assets.
\end{IEEEbiography}
\jvspace{-3em}
\begin{IEEEbiography}[{\includegraphics[width=1in,height=1.25in,clip,keepaspectratio]{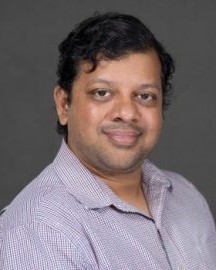}}]{Ram Rajagopal} is an Associate Professor of Electrical Engineering as well as Civil and Environmental Engineering at Stanford University, where he directs the Sustainable Systems Lab, focused on large-scale monitoring, data analytics, and stochastic control for infrastructure networks, in particular, power networks. He received the Ph.D. degree in Electrical Engineering and Computer Sciences from the University of California, Berkeley. His current research interests are in the integration of renewables, smart distribution systems, and demand-side data analytics.
\end{IEEEbiography}
\jvspace{-3em}
\begin{IEEEbiography}[{\includegraphics[width=1in,height=1.25in,clip,keepaspectratio]{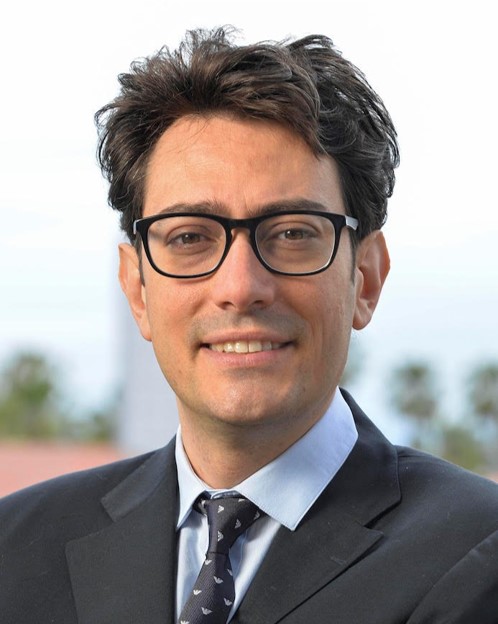}}]{Marco Pavone} is an Associate Professor of Aeronautics and Astronautics at Stanford University, where he is the Director of the Autonomous Systems Laboratory.
He received a Ph.D. degree in Aeronautics and Astronautics from MIT in 2010. His main research interests are in the development of methodologies for the analysis, design, and control of autonomous systems, with an emphasis on self-driving cars, autonomous aerospace vehicles, and future mobility systems. 
He is currently  an Associate Editor for the IEEE Control Systems Magazine.
\end{IEEEbiography}

\end{document}